\documentclass[reprint,pra,notitlepage,floatfix,nofootinbib]{revtex4-1}
\usepackage{newtxtext}
\usepackage{newtxmath}
\usepackage{bbm}
\let\lambda\lambdaup

\usepackage{amsmath}
\usepackage{amssymb}
\allowdisplaybreaks
\usepackage{graphicx}
\usepackage[x11names]{xcolor}
\usepackage[colorlinks=true,linkcolor=blue,citecolor=blue,urlcolor=blue]{hyperref}
\graphicspath{{fig/}}

\renewcommand{\v}[1]{\mathbf{#1}}
\newcommand{\vv}[1]{\boldsymbol{#1}}

\renewcommand{\Re}{\mathop{\mathrm{Re}}}

\DeclareMathOperator{\cov}{cov}
\DeclareMathOperator{\tr}{tr}

\begin{document}

\title{Neural network wave functions and the sign problem}
\author{Attila Szab\'o}
\author{Claudio Castelnovo}
\affiliation{TCM Group, Cavendish Laboratory, University of Cambridge, Cambridge CB3 0HE, United Kingdom}

\begin{abstract}
    Neural quantum states (NQS) are a promising approach to study many-body quantum physics. However, they face a major challenge when applied to lattice models: Convolutional networks struggle to converge to ground states with a nontrivial sign structure.
    We tackle this problem by proposing a neural network architecture with a simple, explicit, and interpretable phase ansatz, which can robustly represent such states and achieve state-of-the-art variational energies for both conventional and frustrated antiferromagnets. 
    In the latter case, our approach uncovers low-energy states that exhibit the Marshall sign rule and are therefore inconsistent with the expected ground state.
    Such states are the likely cause of the obstruction for NQS-based variational Monte Carlo to access the true ground states of these systems. 
    We discuss the implications of this observation and suggest potential strategies to overcome the problem.
\end{abstract}

\maketitle
%
%
%%%%%%%%%%%%%%%%%%%%%%%%%%%%%%%%%%%%%%%%%%%%%%%%%%%%%%%%%%%%%%

\section{Introduction}
Over the last decade, machine learning has had a profound impact on nearly all aspects of life as well as the physical sciences~\cite{Carleo2019MachineSciences}.
In condensed matter physics in particular, using neural networks as ans\"atze for quantum many-body wave functions has emerged as an exciting application of machine learning techniques to solve challenging problems. 
Since the first demonstration~\cite{Carleo2017SolvingNetworks} of the restricted Boltzmann machine (RBM)~\cite{Melko2019RestrictedPhysics} as a practical wave function ansatz for obtaining ground states of many-body Hamiltonians using variational Monte Carlo (VMC) techniques,
such neural quantum states (NQS), including deep convolutional networks~\cite{Choo2019StudyStates,Choo2020FermionicStructure,Sharir2020DeepSystems,Yang2020DeepPhysics}, have become an important branch of many-body numerical techniques, competitive with, and sometimes even outperforming, state-of-the-art tensor network (TN) methods. 
NQS approaches are fundamentally appealing because both RBMs~\cite{LeRoux2008RepresentationalNetworks} and deep neural networks~\cite{Cybenko1989ApproximationFunction} can represent almost any function accurately, without a priori limitations like the Monte Carlo sign problem~\cite{Li2019Sign-Problem-FreeApplications} or the area law entanglement of TNs~\cite{Wolf2008AreaCorrelations,*Riera2006AreaNetworks,Orus2019TensorSystems,Levine2019QuantumArchitectures}. 
RBM states have also been used successfully in higher dimensions~\cite{Carleo2017SolvingNetworks,Choo2019StudyStates} and for chiral topological states~\cite{Kaubruegger2018ChiralNetworks}, both of which pose well-known difficulties to TN techniques. 

All promising properties notwithstanding, NQS approaches are not without their own difficulties. 
In particular, while NQS ans\"atze can in principle represent nontrivial sign structures, actually learning them appears to pose significant challenges, especially in frustrated systems.
This appears less pointedly in RBMs and other shallow, fully connected architectures, which are able to reach low variational energies even for Hamiltonians with a severe sign problem~\cite{Choo2018SymmetriesStates,Cai2018ApproximatingNetworks,Hendry2019MachineSystems,Torlai2019WavefunctionDifferentiation,Westerhout2020GeneralizationStates,Melko2019RestrictedPhysics,Nomura2020MachineCalculations}.
By contrast, deep convolutional networks (which are desirable for cutting-edge applications due to their better scalability and explicit translational invariance~\cite{dAscoli2019FindingBias}) quite often fail to converge to ground states with near-zero average signs (e.g., of antiferromagnetic or fermionic systems)~\cite{Choo2019StudyStates}. 
Attempting to learn such states can generate unphysically rough amplitude profiles, resulting in poor convergence or even complete breakdown of the protocol.
Successful variational learning of NQS ground states in these cases requires transforming the Hamiltonian to remove its sign problem, severely limiting the usefulness of the method~\cite{Choo2019StudyStates,Li2019Sign-Problem-FreeApplications}.%
\footnote{For instance, the sign problem of unfrustrated antiferromagnets can be cured by imposing the Marshall sign rule~\cite{Marshall1955Antiferromagnetism}; the same was used successfully to stabilise the learning of frustrated ground states~\cite{Choo2019StudyStates}. To the best of our knowledge, the only convolutional NQS that successfully approached an antiferromagnetic ground state without such preconditioning is that of {Ref.~\onlinecite{Liang2018SolvingNetworks}}, which, however, produces variational energies quite far from the state of the art even with a large number of adjustable parameters.}

The origins of this ``sign problem'' remain poorly understood, and its existence is counterintuitive given the success of convolutional networks in a range of machine learning applications~\cite{Goodfellow2016DeepLearning}. 
Some insight has recently been offered in Ref.~\onlinecite{Westerhout2020GeneralizationStates}, which studied the ability of NQS ans\"atze to reconstruct the exact ground state from partial data in a supervised learning scenario.
In certain frustrated phases, a range of architectures show poor generalisation properties, especially when it comes to representing their highly nontrivial sign structures. 
This also impedes the convergence of VMC algorithms, which rely on reconstructing quantum expectation values based on a small sample of the Hilbert space~\cite{Becca2017QuantumSystems}.
Further work to understand these phenomena and develop more robust NQS ans\"atze is therefore crucial to deploy neural network-based VMC algorithms to study many interesting and challenging problems in condensed matter physics.

In this paper, we make a contribution to this quest by introducing a NQS architecture, and a corresponding variational optimisation protocol, which can reliably find low-energy variational states of antiferromagnetic Hamiltonians without any prior knowledge of the sign structure of the ground state wave function. 
Our ansatz has a simple, explicit, and interpretable phase representation, whose convergence properties improve even upon deep convolutional networks. 
We benchmark our approach on the spin-1/2 $J_1$--$J_2$ Heisenberg antiferromagnet (HAFM) on the square lattice, and achieve variational energies comparable to the state of the art~\cite{Hu2013DirectAntiferromagnetism,Gong2014PlaquetteModel,Choo2019StudyStates} both at the unfrustrated point $J_2=0$ and in the fully frustrated quantum spin liquid phase at $J_2/J_1=0.5$. 

In the unfrustrated case, our approach is able to learn the expected Marshall sign rule (MSR) with excellent accuracy.
This is a crucial improvement over previous VMC protocols based on convolutional NQS, which suffer from the sign problem even in this simpler case~\cite{Choo2019StudyStates}. 
Our approach, therefore, paves the way for systematically studying conventional phases, including critical ones~\cite{Zen2020FindingStates}, where the area law entanglement of TNs is a serious impediment~\cite{Orus2019TensorSystems}.

In the frustrated case, we again achieve an excellent variational energy; however, we find a state that obeys the same MSR, even though the true ground state is expected to deviate from it significantly. 
The existence of such ``MSR-like'' low-energy variational states, and the ease and stability with which the VMC algorithm homes in on them,
highlight the risks of using the energy as the only criterion for assessing the accuracy of variational wave functions
and may explain the poor generalisation properties of supervised NQS learning in frustrated regimes~\cite{Westerhout2020GeneralizationStates,Cai2018ApproximatingNetworks}.
We suggest potential improvements towards the end of this work.%
\footnote{Beyond NQS, Ref.~\onlinecite{Thibaut2019Long-rangeLattice} proposed a new ``long-range entangled-plaquette state'' variational ansatz that appears to achieve both excellent variational energies and the correct phase structure in strongly frustrated one-dimensional systems. 
It will be interesting to see in the future whether this method remains successful for two-dimensional spin liquids (e.g., the model considered in our work) as well, and what insights its structure may provide for designing better NQS architectures.}
%
%
%%%%%%%%%%%%%%%%%%%%%%%%%%%%%%%%%%%%%%%%%%%%%%%%%%%%%%%%%%%%%%

\section{Our approach} 
We start by separating the amplitudes and phases of the wave function into two neural networks.%
\footnote{We note that this idea is in line with the observation made in Ref.~\onlinecite{Westerhout2020GeneralizationStates} that the amplitudes and phases of the wave function coefficients in NQS have very different generalisation properties, with the former being far easier to generalise than the latter.}
For convenience~\cite{Carleo2019NetKet:Systems,Choo2019StudyStates,Becca2017QuantumSystems}, we take the networks to represent the logarithm of the wave function $|\psi\rangle$:
\begin{equation}
    \log \langle\vv\sigma | \psi\rangle = A(\vv\sigma) + i\Phi(\vv\sigma),
\end{equation}
where $|\vv\sigma\rangle$ are $\sigma^z$ basis states, and $A(\vv\sigma)$ and $\Phi(\vv\sigma)$ are two functions (represented as real-valued neural networks) mapping the basis state to the log-modulus and phase of its complex amplitude, respectively. 

This ansatz is then optimised in two stages. 
First, the phases are optimised to minimise the variational energy while keeping the amplitudes of all $\sigma^z$ basis states equal. 
This allows moving away from the initial guess for $\Phi$ (which resembles the ground state of a \textit{ferromagnetic} Hamiltonian) and approaching the correct phases without scrambling the corresponding amplitude profile. 
Optimal sign structures depend weakly on the imposed amplitudes, a prime example being the MSR for antiferromagnets on bipartite lattices~\cite{Marshall1955Antiferromagnetism}, which minimises the variational energy for \textit{any} given set of amplitudes. 
Therefore, a well-converged result of this first stage is expected to be a good initial guess in the second one, where $A$ and $\Phi$ are optimised simultaneously to approach the true ground state. 

\begin{figure}
    \centering
    \includegraphics{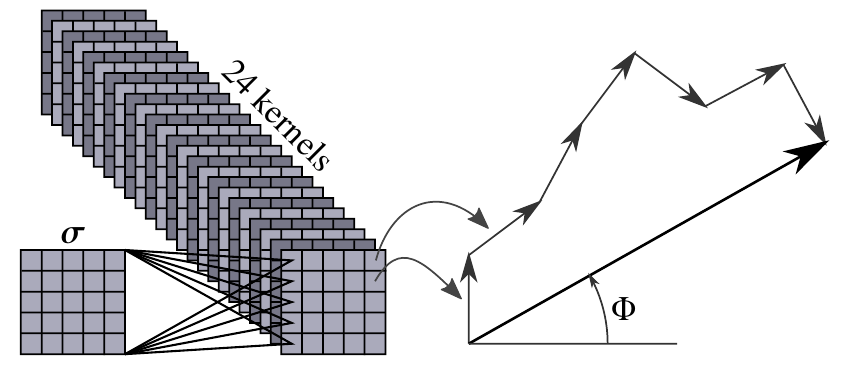}
    \caption{Single-layer convolutional network used to represent the phase $\Phi$ of the wave function. The spins $\vv\sigma$ are mapped through convolutional kernels spanning the entire lattice (with periodic boundary conditions). Each entry in the images is then taken as the argument of a unit complex number; $\Phi$ is given by the argument of their sum. }
    \label{fig: phase machine}
\end{figure}

To make use of this protocol, however, the phases $\Phi$ have to be represented in a way that can approach the true ground state sign structure starting from an initial guess very far from it. 
We propose a single-layer convolutional architecture, visualised in Fig.~\ref{fig: phase machine}, where the activation layer is replaced by summing the convolutional output as phasors:
\begin{equation}
    \Phi(\vv\sigma) = \arg\bigg[\sum_{n,\v r} \exp\Big( i b_n +  \sum_{\v r'} iw_{n,\v r-\v r'}\sigma_{\v r'}^z \Big)\bigg],
    \label{eq: phase machine}
\end{equation}
where $w_{n,\v r}$ and $b_n$ are the real-valued weights and uniform biases of the convolutional filters, respectively.

Once an appropriate phase structure is found, good variational energies can readily be obtained using any typical neural network architecture to represent $A(\vv\sigma)$. Similar to most machine learning tasks~\cite{Goodfellow2016DeepLearning}, we found that deeper, wider networks generally perform better. In our numerical experiments, a six-layer convolutional network was used, details of which are given in Appendix~\ref{Ssec: architecture}.

We performed the optimisation using stochastic reconfiguration (SR), which approximates the imaginary time evolution of the initial state~\cite{Becca2017QuantumSystems}. For neural network quantum states, this protocol has been shown to be superior to stochastic gradient descent and other commonly used algorithms in deep learning~\cite{Park2020GeometryStates}. SR is described in Appendix~\ref{Ssec: stochastic reconfig} and details of the optimisation are given in Appendix~\ref{Ssec: protocol}. The simulations were implemented using the NetKet library~\cite{Carleo2019NetKet:Systems}.
%
%
%%%%%%%%%%%%%%%%%%%%%%%%%%%%%%%%%%%%%%%%%%%%%%%%%%%%%%%%%%%%%%

\section{Numerical experiments}
We deployed our method to the spin-1/2 HAFM on the square lattice with nearest and next-nearest neighbour interactions:
\begin{equation}
    H = J_1 \sum_{\langle ij\rangle} \vec\sigma_i \cdot\vec\sigma_j + J_2\sum_{\langle\langle ij\rangle\rangle}  \vec\sigma_i \cdot\vec\sigma_j,
    \label{eq: hamiltonian}
\end{equation}
where $J_1,J_2\ge0$ and $\langle ij\rangle$ and $\langle\langle ij\rangle\rangle$ refer to nearest and second neighbour sites, respectively. We considered a $10\times10$ lattice with periodic boundary conditions and set $J_1=1$ without loss of generality.

\begin{figure}
    \centering
    \includegraphics{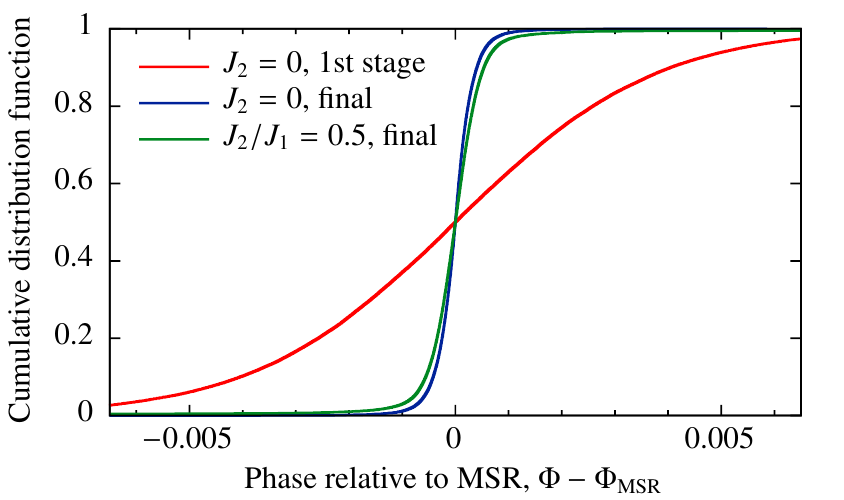}
    \caption{Cumulative distribution function of Marshall-adjusted phases $\Phi-\Phi_\mathrm{MSR}$ for the unfrustrated case $J_2=0$, learned before (red) and after (blue) optimising the amplitudes, and for the final state of the frustrated case $J_2/J_1=0.5$ (green). All distributions  are very sharply peaked, indicating that $\Phi$ is a good approximation of the MSR~\eqref{eq: Marshall sign rule}. The standard deviations of the distributions are, in order, $3.3\times 10^{-3}$, $4.0\times 10^{-4}$, and $2.8\times 10^{-3}$.}
    \label{fig: all phases}
\end{figure}

Our first benchmark was the nearest-neighbour Hamiltonian $J_2=0$. In this case, the sign problem can be cured by rotating all spins on a chequerboard sublattice $A$ of the square lattice by $\pi$ around the $\sigma^z$ axis:
\begin{align}
    \sigma^x &\to -\sigma^x, &
    \sigma^y &\to -\sigma^y, &
    \sigma^z &\to \sigma^z,
    \label{eq: Marshall spin transform}
\end{align}
as this makes the coefficients of all off-diagonal terms $\sigma^+_i\sigma^-_j$ negative.
As a result, $\langle\vv\sigma|\prod_{i\in A} \sigma_i^z |\mathrm{GS}\rangle$ is positive (up to an overall phase) for all $\sigma^z$ basis states $|\vv\sigma\rangle$. In terms of the phase structure $\Phi(\vv\sigma)$, the resulting Marshall sign rule (MSR) can be written as
\begin{equation}
    \Phi_\mathrm{MSR}(\vv\sigma) = \pi \sum_{i\in A} \frac{1-\sigma_i^z}2.
    \label{eq: Marshall sign rule}
\end{equation}
We find that our phase structure ansatz~\eqref{eq: phase machine} converges reliably to~\eqref{eq: Marshall sign rule} in the first stage of the optimisation, as shown in Fig.~\ref{fig: all phases}. 
This is to be expected as the same sign structure attains optimal variational energy for any set of amplitudes~\cite{Marshall1955Antiferromagnetism}, including the  uniform one used here.
Other convolutional networks, by contrast, fail to approach the MSR, which in turn leads to instabilities in the amplitude optimisation, as shown in Appendix~\ref{Ssec: phase structures}.

In the subsequent optimisation of amplitudes and phases, we achieved a variational energy of $-0.671275$ per spin, $2.7\times 10^{-4}$ higher than the numerically exact energy given by stochastic series expansion~\cite{Mezzacapo2009Ground-stateCarlo}.
This energy is only slightly above the one attained by the convolutional network of Ref.~\cite{Choo2019StudyStates}, even though the latter has substantially more variational parameters (7676 compared to our 5145) and is preconditioned with the exact MSR. 

We then used our approach to study the fully frustrated phase of the model at $J_2/J_1=0.5$~\cite{Hu2013DirectAntiferromagnetism,Gong2014PlaquetteModel}. We achieved a variational energy of $-0.494757$ per spin, $2.8\times 10^{-3}$ higher than the best energies obtained by Lanczos iterating a Gutzwiller projected fermionic wave function~\cite{Hu2013DirectAntiferromagnetism}. Our result again compares favourably with the best NQS-based variational energy in the literature~\cite{Ferrari2019NeuralFunctions}, where the corresponding error is $1.8\times 10^{-3}$.
The variational energies obtained in both cases, together with relevant benchmarks, are summarised in Table~\ref{tab: variational energies}.

\begin{table}
    \centering
    \setlength{\tabcolsep}{0.75em}
    \begin{tabular}{rll}
            & \multicolumn{1}{c}{$J_2=0$} 
                                & \multicolumn{1}{c}{$J_2/J_1=0.5$} \\\hline
        our work
            & $-0.671275(5)$    & $-0.494757(12)$   \\\hline
        GWF~\cite{Hu2013DirectAntiferromagnetism} 
            & $-0.66935(1)$     & $-0.49439(1)$     \\
        GWF\,+\,RBM~\cite{Ferrari2019NeuralFunctions}
            & $-0.67111(2)$     & $-0.49575(3)$     \\
        CNN~\cite{Choo2019StudyStates} 
            & $-0.67135(1)$     & $-0.49516(1)$     \\
        best~\cite{Mezzacapo2009Ground-stateCarlo,Hu2013DirectAntiferromagnetism}
            & $-0.671549(4)$    & $-0.49755(1)$
    \end{tabular}
    \caption{Variational energies (in units of $J_1$ per spin) attained in this work compared to other state-of-the-art energies on the same system. Our approach consistently outperforms plain Gutzwiller projected fermionic wave functions (GWF)~\cite{Hu2013DirectAntiferromagnetism}, and achieves similar accuracy to the RBM-enhanced GWF of Ref.~\onlinecite{Ferrari2019NeuralFunctions} and the convolutional networks of Ref.~\onlinecite{Choo2019StudyStates} (CNN). The ``best'' energy is obtained using numerically exact stochastic series expansion for $J_2=0$~\cite{Mezzacapo2009Ground-stateCarlo} and Lanczos-corrected GWF for $J_2/J_1=0.5$~\cite{Hu2013DirectAntiferromagnetism}.}
    \label{tab: variational energies}
\end{table}

Surprisingly, however, we find that the converged phase structure $\Phi(\vv\sigma)$ in the frustrated case recovers the MSR to a high accuracy, and no bimodality consistent with having both positive and negative amplitudes can be seen (see Fig.~\ref{fig: all phases}). This is at odds with the fact that the frustrated Hamiltonian~\eqref{eq: hamiltonian} remains non-stoquastic even after the Marshall transformation~\eqref{eq: Marshall spin transform}, and as such its average sign should fall below 1~\cite{Troyer2005ComputationalSimulations}. However, the MSR is expected to remain a relatively good approximation of the exact ground state sign structure (see Appendix~\ref{Ssec: average sign}) and for smaller but finite values of $J_2$, it remains exact~\cite{Richter1994OnAntiferromagnet,*Voigt1997Marshall-PeierlsAntiferromagnet}.

To further quantify the quality of the variational wave functions given by our approach, we evaluated their total spin $\langle \vec S^2\rangle$, the expectation value of the parity operator $\mathcal{P} = \prod \sigma^x$, as well as the statistical weight of the five irreducible representations (irreps) of the point group $D_4$ of the square lattice in the wave function. 
Since $\vec S^2$, $\mathcal{P}$, and the point group symmetry operators all commute with the Hamiltonian~\eqref{eq: hamiltonian}, the true ground state is an eigenstate of the former two, and transforms according to precisely one irrep of the latter; 
furthermore, ground states of HAFMs are normally singlets ($\vec S^2=0$) and thus have even parity:
Deviations from these expectations can be used as a quantitative test of the converged wave functions.
These results are shown in Table~\ref{tab: observables}; computational details are given in Appendix~\ref{Ssec: estimation}.
We achieve similar figures of $\langle \vec S^2\rangle$ to those of Ref.~\onlinecite{Choo2019StudyStates} in both cases, which, together with the very low weight of parity odd states, suggests a small admixture of states with high spin quantum numbers.%
\footnote{$\protect\langle \vec S^2\protect\rangle\approx0.6$  in the frustrated case could be consistent with a 70--30 mixture of singlet and triplet states; however, this would yield an average parity of $0.4$. While one cannot rule out a large $s=2$ admixture on these grounds, it is more natural to assume contributions with a range of higher spin quantum numbers.}
We also note that discrepancies from the ideal ground state are an order of magnitude larger in the frustrated case by all three measures: this is consistent with the worse energy convergence and sign structure discrepancies of the same.
Finally, we evaluated the antiferromagnetic order parameter
\begin{equation}
    S^2(\vec q) = \frac1{N(N+2)} \sum_{i,j} \langle \vec \sigma_i\cdot \vec \sigma_j\rangle e^{i\vec q\cdot (\vec r_i-\vec r_j) }
    \label{eq: order parameters}
\end{equation}
for $\vec q=(\pi,0)$ and $(\pi,\pi)$, which correspond to stripy and N\'eel orders, respectively: The results are consistent with those plotted in Ref.~\onlinecite{Choo2019StudyStates}.

\begin{table}
    \centering
    \setlength{\tabcolsep}{0.75em}
    \begin{tabular}{rll}
            & \multicolumn{1}{c}{$J_2=0$} 
                                & \multicolumn{1}{c}{$J_2/J_1=0.5$} \\\hline
        Parity $\langle \mathcal{P}\rangle$ 
            &     0.998373(29)  &     0.990426(69)  \\\hline
        Weight of irrep $\boldsymbol{A_1}$
            & \bf 0.998645(18)  & \bf 0.989363(51)  \\
        $A_2$ 
            &     0.000142(6)   &     0.000928(15)  \\
        $B_1$
            &     0.000283(8)   &     0.003335(29)  \\
        $B_2$ 
            &     0.000167(6)   &     0.001169(17)  \\
        $E$   
            &     0.000763(14)  &     0.005205(36)  \\\hline
        Total spin $\langle \vec S^2\rangle$
            &     0.065(21)     &     0.581(43)     \\\hline
        Stripy o.p.\ $S^2(\pi,0)$
            &     0.00498(5)    &     0.00521(7)    \\
        N\'eel o.p.\ $S^2(\pi,\pi)$
            &     0.1571(2)     &     0.0633(2)     
    \end{tabular}
    \caption{Average parity $\langle\mathcal{P}\rangle$, total spin $\langle \vec S^2\rangle$, and antiferromagnetic order parameters of, and statistical weight of irreps of the point group $D_4$ in, the fully converged NQS wave functions in the unfrustrated limit $J_2=0$ and for $J_2/J_1=0.5$. Both wave functions are predominantly parity even and transform according to the trivial representation $A_1$ (bold); the weights of states with odd parity and/or different spatial symmetry are $\approx0.001$ and $\approx0.01$ in the two cases. The converged $\langle \vec S^2\rangle$ is similarly larger in the frustrated case, consistent with its worse energy convergence.}
    \label{tab: observables}
\end{table}
%
%
%%%%%%%%%%%%%%%%%%%%%%%%%%%%%%%%%%%%%%%%%%%%%%%%%%%%%%%%%%%%%%

\section{Discussion}
\label{sec: discussion}

We developed a robust and efficient protocol for finding low energy states with a nontrivial sign structure using convolutional neural quantum states without any prior knowledge on the sign problem of the Hamiltonian. 
We used an ansatz with two neural networks that represent the amplitudes and phases separately, and optimised it in two stages, first generating an approximate phase structure, from which the entire wave function can readily converge without encountering severe instabilities.
We demonstrated our approach by attempting to learn the ground states of the square lattice spin-1/2  $J_1$--$J_2$ HAFM both at the unfrustrated point $J_2=0$ and at $J_2/J_1=0.5$, inside the fully frustrated spin liquid phase.
In both cases, we reached variational energies comparable to the best NQS energies reported in the literature~\cite{Choo2019StudyStates}; the difference might be attributed to the smaller number of variational parameters in our ansatz. 
Importantly, we used a fully convolutional architecture: This automatically imposes translational invariance, a useful inductive bias that speeds up the convergence to a robust state representation~\cite{dAscoli2019FindingBias} and allows for resolving the lowest energy states in different symmetry sectors~\cite{Choo2018SymmetriesStates}. 
Furthermore, the convolutional structure reduces the number of variational parameters from the $O(N^2)$ typical for RBMs and other fully connected architectures~\cite{Sehayek2019LearnabilityMachines,Nomura2020MachineCalculations} to $O(N)$, which keeps VMC algorithms viable for larger system sizes.

\begin{figure}
    \centering
    \includegraphics{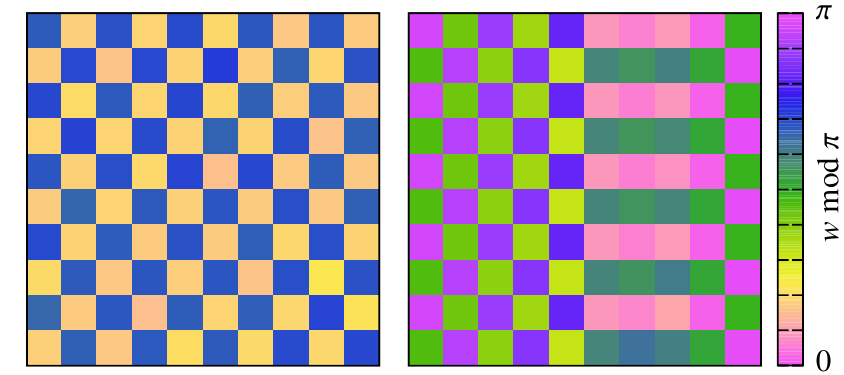}
    \caption{Weights $w_{\v r}$ of a typical convolutional kernel converged for $J_2=0$ (left) and $J_2/J_1=0.5$ (right). The chequerboard pattern of the former is a direct consequence of the Marshall sign rule (see Appendix~\ref{Ssec: kernels}); the latter shows an admixture of a stripy antiferromagnetic pattern, consistent with the MSR for the opposite limit, $J_1=0$.
    (Perceptionally uniform colour map chosen following Ref.~\onlinecite{Kovesi2015GoodThem}.)}
    \label{fig: kernels}
\end{figure}

At $J_2=0$, our phase structure ansatz~\eqref{eq: phase machine} learns the Marshall sign rule with better generalisation properties than other convolutional networks, both deep and shallow, which is crucial for finding ground states reliably~\cite{Westerhout2020GeneralizationStates}. 
A possible origin of the underlying inductive bias is the following:
$\sigma_i^+\sigma_j^-$ exchanges a positive and a negative value of $\sigma_{\v r}$ in~\eqref{fig: phase machine}, leading to each phasor changing its phase by $\Delta\phi_{n,\v r} = 2(w_{n,\v r-\v r_i}-w_{n,\v r-\v r_j})$, where the dummy variable $\v r$ covers the entire lattice. 
The change of the overall phase $\Phi$ is an ``average'' of these. 
While the energy of an antiferromagnetic interaction is optimised if $\Delta\Phi=\pi$, this can be realised by a range of distributions of the $\Delta\phi$ centred on $\pi$, suggesting that the MSR can be encoded in a robust way by such an architecture.
Indeed, the weights $w_{\v r}$ produced by VMC in the unfrustrated case show a distinct chequerboard pattern that produces $\Delta\phi\approx\pi$ for all nearest neighbour pairs, consistent with the MSR (see Fig.~\ref{fig: kernels}).
By contrast, deep neural networks have an inductive bias for functions that only change significantly upon large-scale changes of the input~\cite{Lin2017WhyWell,*Valle-Perez2018DeepFunctions,*DePalma2018RandomFunctions,Goodfellow2016DeepLearning}.
While this is desirable for most machine learning applications, it is detrimental for learning a nontrivial quantum phase structure. 

In the frustrated case, the same approach fails to find the appropriate ground state sign structure, homing in instead on the MSR. 
The existence of low-lying variational states with ``simple'' sign structures deep within frustrated phases parallels the poor generalisation of the corresponding ground states in supervised learning scenarios~\cite{Westerhout2020GeneralizationStates}, hinting at a possible bias of NQS ans\"atze towards such states and the corresponding ubiquity of this ``residual sign problem''.
This issue is compounded with the gradual nature of VMC optimisation, which makes it prone to get stuck in such minima. By contrast, tensor network wave functions are typically optimised using singular value decompositions~\cite{Orus2019TensorSystems}, which allows large yet controlled changes in the wave function, yielding high-quality ground state estimates for a range of challenging frustrated systems~\cite{Schollwock2005TheGroup}. Developing similar methods for NQS states may thus help overcome this problem. 

Beyond the explicit sign structure, the converged variational state in the frustrated case obeys the spin rotation and spatial symmetries of the Hamiltonian~\eqref{eq: hamiltonian} substantially less accurately than for the nearest-neighbour model: 
$\langle \vec S^2\rangle$ as well as the statistical weights of parity odd states and those that do not transform according to the dominant irrep of the point group are all an order of magnitude higher for the former.
While this may be a straightforward consequence of the gapless spin liquid ground state~\cite{Hu2013DirectAntiferromagnetism}, it might also be necessary to reconcile low variational energies with the MSR and serve as a signature of the residual sign problem.
If this is the case, the performance of our approach might be improved substantially by imposing symmetries, either through making the wave function explicitly symmetric~\cite{Choo2019StudyStates,Choo2018SymmetriesStates,Nomura2020MachineCalculations,Vieijra2020RestrictedSymmetries}, or using the variational protocol to project out states that do not have the right symmetry (e.g., ones for which $\langle \vec{S}^2\rangle\neq0$). 

We also note that the representation of the MSR learned in the two cases is starkly different (see Fig.~\ref{fig: kernels} and Appendix~\ref{Ssec: kernels} for a detailed analysis):
While the unfrustrated sign structure attains the simplest possible representation of the MSR, 
stripe features consistent with the MSR of the opposite unfrustrated limit, $J_1=0$, appear in the frustrated case.
This hints at an (ultimately failed) attempt at learning a ``compromise'' between the two limits, consistent with our qualitative understanding of frustrated phases. 
More detailed insight into this learning dynamics may open up the possibility of finding similarly simple ans\"atze with significantly better generalisation abilities.

Finally, we believe that the simplicity, interpretability, and robustness of our phase representation, as well as the insight it affords us about the sign problem of NQS ans\"atze, make it a useful resource to guide efforts to design novel network architectures and training protocols that will one day reliably learn frustrated ground states with complex phase structures. 
Successfully learning the MSR also suggests that our method can readily be used for large-scale simulations of conventional phases, including excited states~\cite{Choo2018SymmetriesStates,Nomura2020MachineCalculations,Hendry2019MachineSystems} and gapless or critical systems~\cite{Zen2020FindingStates}, without the entanglement limitations of tensor networks~\cite{Levine2019QuantumArchitectures}.

%
%
%%%%%%%%%%%%%%%%%%%%%%%%%%%%%%%%%%%%%%%%%%%%%%%%%%%%%%%%%%%%%%

\section*{Acknowledgments}
We thank Kenny Choo, Adrian Feiguin, and Titus Neupert for illuminating discussions, and Nigel Cooper and Ben Simons for sharing computational resources with us. This work was supported in part by the Engineering and Physical Sciences Research Council (EPSRC) Grants No. EP/M007065/1 and No. EP/P034616/1.
%
%
%%%%%%%%%%%%%%%%%%%%%%%%%%%%%%%%%%%%%%%%%%%%%%%%%%%%%%%%%%%%%%

\appendix
\section{Stochastic reconfiguration}
\label{Ssec: stochastic reconfig}

Stochastic reconfiguration (SR) is a generic method for optimising parametrised trial wave functions $|\Psi(\{\alpha_k\})\rangle$ so as to minimise their energy with respect to a Hamiltonian $H$~\cite{Becca2017QuantumSystems,Sorella2001GeneralizedCarlo}.
It is commonly used in variational Monte Carlo studies owing to its more reliable convergence compared to other common optimisation protocols, such as stochastic gradient descent~~\cite{Sorella2001GeneralizedCarlo,Park2020GeometryStates}.

The method proceeds by approximating the imaginary time evolution of the trial wave function using Monte Carlo sampling. Namely, given $|\Psi(\{\alpha_k\})\rangle$, we want to find a new set of the real parameters\footnote{Equivalent expressions can be derived for trial wave functions that are (piecewise) analytic functions of complex parameters~\cite{Becca2017QuantumSystems,Carleo2017SolvingNetworks}. The result is identical to~\eqref{Seq: SR for MC}, omitting the real-part signs.} $\alpha'_k = \alpha_k + \delta\alpha_k$ such that $|\Psi'\rangle = |\Psi(\{\alpha'_k\})\rangle$ is a good approximation to 
\begin{equation}
    |\Psi'_\mathrm{exact}\rangle = e^{-\eta H}|\Psi(\{\alpha_k\})\rangle \approx (1-\eta H)|\Psi(\{\alpha_k\})\rangle,
    \label{Seq: imaginary time evolution}
\end{equation}
where $\eta$ is a small positive number playing the role of the learning rate in machine learning language. 
Since we only want to project out all excited states, the Trotterisation error in~\eqref{Seq: imaginary time evolution} is irrelevant.
%the Trotterisation error is unimportant, since the approximate imaginary time evolution~\eqref{Seq: imaginary time evolution} still projects out all excited states. 
$|\Psi'\rangle$ is optimised by maximising the overlap of the (unnormalised) wave functions $|\Psi'_\mathrm{exact}\rangle$ and $|\Psi'\rangle$:
\begin{equation}
    |C|^2 = \frac{\langle \Psi'_\mathrm{exact}|\Psi'\rangle \langle \Psi' | \Psi'_\mathrm{exact}\rangle}{\langle \Psi'_\mathrm{exact}|\Psi'_\mathrm{exact}\rangle \langle \Psi' | \Psi'\rangle}.
\end{equation}
To linear order in both $\eta$ and $\delta\alpha_k$, the condition $\partial_{\alpha_k} |C|^2=0$ leads to
\begin{align}
    \sum_{j}{}& \delta\alpha_{j} \Re\left[
    \frac{\left\langle \partial_{\alpha_j}\Psi \middle| \partial_{\alpha_k}\Psi\right\rangle }{\langle \Psi | \Psi\rangle } - 
    \frac{\left\langle \partial_{\alpha_j}\Psi \middle|\Psi\right\rangle }{\langle \Psi | \Psi\rangle }
    \frac{\left\langle \Psi \middle| \partial_{\alpha_k}\Psi\right\rangle }{\langle \Psi | \Psi\rangle}
    \right] \nonumber\\
    &= \eta \Re\left[
    \frac{\left\langle \Psi \middle| H \middle| \partial_{\alpha_k}\Psi\right\rangle }{\langle \Psi | \Psi\rangle } 
    - \frac{\langle \Psi |H|\Psi\rangle }{\langle \Psi | \Psi\rangle }
    \frac{\left\langle \Psi \middle| \partial_{\alpha_k}\Psi\right\rangle }{\langle \Psi | \Psi\rangle}
    \right].
    \label{Seq: SR with wave function}
\end{align}
In order to find $\delta\alpha_k$ numerically, we want to rewrite the expectation values in~\eqref{Seq: SR with wave function} as Monte Carlo averages with respect to the quantum probability distribution $p(\vv\sigma) = \big|\langle \vv\sigma | \Psi \rangle\big|^2/ \langle \Psi |\Psi\rangle$. This can readily be done by inserting a resolution of the identity; for example, we have
\begin{align}
    \frac{\left\langle \Psi \middle| H \middle| \partial_{\alpha_k}\Psi\right\rangle }{\langle \Psi | \Psi\rangle } &= 
    \sum_{\vv\sigma} \frac{\langle \Psi |H|\vv\sigma\rangle \left\langle\vv\sigma\middle| \partial_{\alpha_k}\Psi\right\rangle }{\langle \Psi | \Psi\rangle } \nonumber\\
    &= \sum_{\vv\sigma} \frac{\big|\langle \vv\sigma | \Psi \rangle\big|^2}{\langle \Psi |\Psi\rangle} \frac{\langle \Psi |H|\vv\sigma\rangle}{\langle \Psi |\vv\sigma\rangle} \frac{\left\langle\vv\sigma\middle| \partial_{\alpha_k}\Psi\right\rangle}{\langle\vv\sigma| \Psi\rangle} \nonumber\\
    &= \sum_{\vv\sigma} p(\vv\sigma) E_\mathrm{loc}^*(\vv\sigma) O_k(\vv\sigma),
    \label{Seq: SR rewriting example}
\end{align}
where we introduce
\begin{equation}
    O_k(\vv\sigma) = \frac{\left\langle\vv\sigma\middle| \partial_{\alpha_k}\Psi\right\rangle}{\langle\vv\sigma| \Psi\rangle} = \partial_{\alpha_k} \log \langle\vv\sigma| \Psi\rangle
\end{equation}
and the local energy
\begin{equation}
    E_\mathrm{loc}(\vv\sigma) = \frac{\langle\vv\sigma | H| \Psi\rangle}{\langle \vv\sigma |\Psi\rangle}.
    \label{Seq: local energy}
\end{equation}
The expectation value~\eqref{Seq: SR rewriting example} can now be estimated as the Monte Carlo average of $E_\mathrm{loc}^* O_k$ for samples distributed according to $p(\vv\sigma)$; the others follow from analogous considerations, resulting in
\begin{equation}
    \sum_{j} \underbrace{\Re \cov( O_{j},  O_k)}_{S_{kj}} \delta\alpha_{j} = -\eta\underbrace{\Re\cov( E_\mathrm{loc}, O_k)}_{f_k}.
    \label{Seq: SR for MC}
\end{equation}
Since the covariance matrix $S$ depends entirely on the parametrisation of the wave function rather than its energy under the Hamiltonian, it can be thought of as a metric tensor on the parametrised Hilbert space~\cite{Becca2017QuantumSystems}. Equation~\eqref{Seq: SR for MC} is thus analogous to the natural gradient approaches used to stabilise gradient descent in other machine learning contexts~\cite{Park2020GeometryStates}.
To improve numerical stability, the covariance is calculated as
\begin{equation*}
    \cov(X,Y) = \langle X^* Y\rangle - \langle X^*\rangle \langle Y\rangle = \langle (X - \langle X\rangle)^*  (Y - \langle Y\rangle)\rangle.
\end{equation*}

In our setup, $\log \langle\vv\sigma| \Psi\rangle=A(\vv\sigma,\{\kappa\})+i\Phi(\vv\sigma,\{\lambda\})$, where both $A$ and $\Phi$ are real-valued functions of real parameters.
It follows that $O_\kappa$ is real for all $\kappa$ and $O_\lambda$ is imaginary for all $\lambda$, and so all entries of the covariance matrix $S$ connecting a $\kappa$ and a $\lambda$ vanish.
This allows us to solve~\eqref{Seq: SR for MC} for the $\delta\kappa$ and the $\delta\lambda$ separately, speeding up the algorithm.

Solving~\eqref{Seq: SR for MC} for $\delta\alpha_k$ requires inverting the covariance matrix $S$, which, while positive semidefinite, tends to be ill-conditioned even for a large number of Monte Carlo samples~\cite{Becca2017QuantumSystems}.
This can be resolved either by using the pseudoinverse, or, more commonly, by adding a small positive constant to the diagonal entries in order to make the matrix invertible. 
In our case, however, the entries of the $S$ matrix corresponding to the parameters of the amplitude and phase have vastly different values (separated by up to eight orders of magnitude). 
To keep the optimisation of both parts viable, we add $10^{-5}$ times the average diagonal entry (i.e., the trace divided by the number of parameters) to both blocks of the matrix:
\begin{equation*}
    \tilde S_\kappa = S_\kappa + 10^{-5}\frac{\tr S_\kappa}{\dim S_\kappa}\mathbbm{1},
\end{equation*}
and likewise for $S_\lambda$.
%
%
%%%%%%%%%%%%%%%%%%%%%%%%%%%%%%%%%%%%%%%%%%%%%%%%%%%%%%%%%%%%%%

\begin{figure*}
    \centering
    \includegraphics{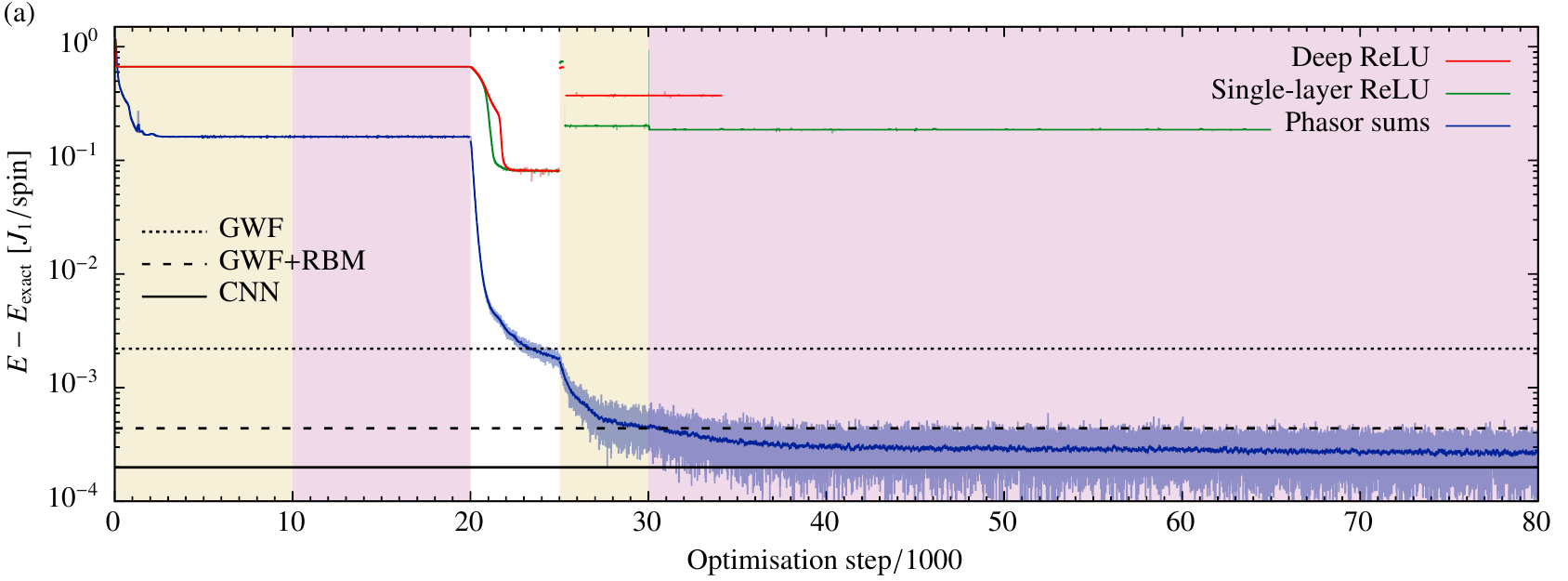}
    \includegraphics{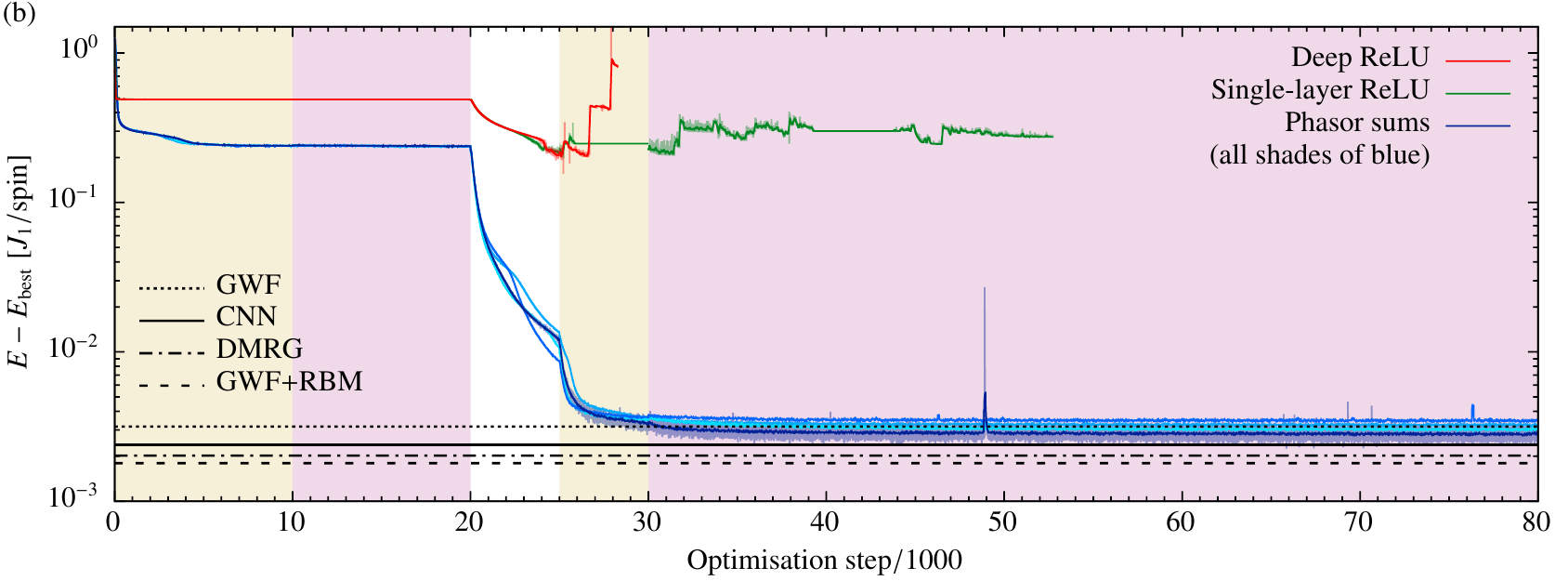}
    \caption{Convergence of our optimisation scheme with various neural network ans\"atze (described in Appendices~\ref{Ssec: architecture} and~\ref{Ssec: phase structures}) for the nearest-neighbour (top panel) and the $J_2/J_1=0.5$ (bottom panel) square lattice HAFM. The shaded area shows the full spread of energy estimates used by the SR algorithm, the thicker lines show 100-step moving averages. In the frustrated case, the best of four runs (dark blue) is shown in detail together with the moving averages of the others (shades of turquoise). The background shading indicates the learning rate $\eta$ (white: 0.001, yellow: 0.01, purple: 0.05). The phasor sum ansatz~\eqref{eq: phase machine} (blue curve) converges reliably to energies close to the true ground state; other ans\"atze (red and green curves; see Appendix~\ref{Ssec: phase structures} for details), both shallow and deep, fail to reach either a consistent variational energy, or one close to the ground state.
    Energies are compared to exact stochastic series expansion for $J_2=0$~\cite{Mezzacapo2009Ground-stateCarlo} or Lanczos-corrected Gutzwiller projected fermionic wave functions (GWF) for $J_2/J_1=0.5$~\cite{Hu2013DirectAntiferromagnetism}. For reference, variational energies are also shown for the convolutional network of Ref.~\onlinecite{Choo2019StudyStates} (CNN), plain~\cite{Hu2013DirectAntiferromagnetism} and RBM-improved~\cite{Ferrari2019NeuralFunctions} GWF, as well as DMRG~\cite{Gong2014PlaquetteModel} for the frustrated case.
    }
    \label{Sfig: evolution}
\end{figure*}

\section{Details of numerical experiments}
\label{Ssec: numerical expt everything}

\subsection{Neural network architectures}
\label{Ssec: architecture}

The amplitude structure $A(\vv\sigma)$ was represented using a six-layer convolutional neural network in all numerical experiments. The layers consist of 8, 7, 6, 5, 4, and 3 $10\times10$ lattice replicas, respectively, which are connected by convolutional filters with real-valued kernels spanning $4\times4$ sites in periodic boundary conditions, and (real-valued) ReLU activation functions:
\begin{equation*}
    f_{1...5}(x) = \left\{\begin{array}{ll}
        x & \quad x\ge 0 \\
        0 & \quad x < 0
    \end{array}\right..
\end{equation*}
The amplitude is given by the modulus of the product of all entries in the last convolutional layer. Since the NQS networks implemented in NetKet represent the logarithm of wave functions~\cite{Carleo2019NetKet:Systems}, this is achieved using a final activation function $f_6(x) = \ln |x|$, followed by summing all entries.
All convolutional layers before the last one are initialised with Gaussian distributed random numbers of zero mean, and standard deviation chosen so as to preserve the typical magnitude of backpropagated derivatives~\cite{He2015DelvingClassification}. The last set of kernels are initialised with a uniform bias of $1.0$ and Gaussian distributed kernel entries with standard deviation $2\times 10^{-4}$. This results in amplitudes uniformly close to 1 upon initialisation.

For all data discussed in the main text, the phases $\Phi(\vv\sigma)$ were represented by the phasor sum ansatz~\eqref{eq: phase machine}. We employed 24 lattice replicas (cf.~Fig.~\ref{fig: phase machine}). Similarly good results are achieved using fewer replicas, but the wider network allows for faster and more reliable convergence.
The convolutional filters are initialised with Gaussian distributed random numbers of standard deviation $0.043$.

\subsection{Optimisation protocol}
\label{Ssec: protocol}

We optimised each wave function ansatz via stochastic reconfiguration in two stages.
First, the phases are optimised with a uniform amplitude distribution (i.e., setting $A\equiv0$): 10\,000 such SR steps with learning rate $\eta=0.01$ were followed by 10\,000 steps with $\eta=0.05$. 
Next, both $A$ and $\Phi$ were optimised starting from the phase distribution achieved in the first stage: for this, we used 5\,000 steps with $\eta=0.001$, 5\,000 steps with $\eta=0.01$, and, finally, 50\,000 steps with $\eta=0.05$. 
The learning rate was increased during the optimisation, because the imaginary time evolution emulated by SR results in infinitesimal temperature (and thus energy) reduction close to the ground state. 
In both stages, the Monte Carlo averages in \eqref{Seq: SR for MC} were evaluated using 5\,000 samples obtained via the Metropolis--Hastings algorithm as implemented by NetKet~\cite{Carleo2019NetKet:Systems}.

The convergence of the phasor sum ansatz to the ground state is shown by the blue curves in Fig.~\ref{Sfig: evolution}. The first stage quickly attains an approximately constant minimum of variational energy. We find, however, that any residual optimisation speeds up the next stage significantly, which is desirable as a single, simple neural network can be evaluated an order of magnitude faster than the full ansatz. The second stage also reaches a nearly converged variational energy in about 20\,000 steps; however, the variational energy is further reduced slightly throughout the procedure.

\subsection{Observable estimation}
\label{Ssec: estimation}

Once the wave function had converged, the estimates of cumulative distribution functions plotted in Figs.~\ref{fig: all phases} and~\ref{Sfig: MSR with bad ansatze} were generated by drawing 100\,000 samples out of the probability distribution $p(\vv\sigma) = \big|\langle \vv\sigma | \Psi \rangle\big|^2/ \langle \Psi |\Psi\rangle$ and sorting their phases.

Analogous to the estimation of variational energies,~\eqref{Seq: local energy}, the expectation value of any operator $A$ can be evaluated as the Monte Carlo average of the local estimates
\begin{equation}
    A_\mathrm{loc}(\vv\sigma) = \frac{\langle\vv\sigma | A | \Psi\rangle}{\langle \vv\sigma |\Psi\rangle}.
    \label{Seq: local operators}
\end{equation}
with respect to the quantum probability distribution $p(\vv\sigma) = \big|\langle \vv\sigma | \Psi \rangle\big|^2/ \langle \Psi |\Psi\rangle$. 
NetKet provides a facility for evaluating such expectation values within the SR protocol: 
This was used to estimate the variational energy, $\langle \vec S^2\rangle$, and the spin correlators~\eqref{eq: order parameters} using 1\,000\,000 Monte Carlo samples. We exploited the translational invariance of the wave function to rewrite the latter two as
\begin{align}
    S^2(\vec q) &= \frac1{N+2} \sum_{i} \langle \vec \sigma_i\cdot \vec \sigma_0\rangle e^{i\vec q\cdot \vec r_i};\\
    \langle \vec S^2\rangle &= N\sum_i \langle \vec\sigma_i\cdot \vec\sigma_0\rangle,
\end{align}
where both sums include $i=0$ (note, however, that $\vec\sigma_0\cdot\vec\sigma_0\equiv 3/4$).

We checked furthermore whether the converged wave functions obey parity and point group symmetries.
The parity operator $\mathcal{P} = \prod_i \sigma_i^x$ commutes with all point group symmetries as well as the Hamiltonian: Therefore, the parity of a wave function is fully characterised by the expectation value $\langle \mathcal{P}\rangle$, without any possibility of symmetry-protected degeneracies.
By contrast, the nonabelian point group $D_4$ gives rise to such degeneracies, limiting the usefulness of plain symmetry operator expectation values. 
Instead, we evaluated the statistical weight of eigenstates transforming according to the different irreps $\alpha$ of $D_4$, using the projection operators~\cite{Heine1960GroupMechanics}
\begin{equation}
    \hat{P}_\alpha = \frac{d_\alpha}{|D_4|}\sum_{g\in D_4} \chi_\alpha(g) \hat{g},
    \label{Seq: irrep projector}
\end{equation}
where $d_\alpha$ and $\chi_\alpha$ are the dimension and characters of the irrep, respectively, and $|D_4|=8$.
The weight of each irrep is given by
\begin{equation*}
    w_\alpha = \frac{\langle \psi|\hat{P}_\alpha^\dagger \hat{P}_\alpha^{\phantom\dagger}|\psi\rangle}{\langle \psi|\psi\rangle} = \frac{\langle \psi|\hat{P}_\alpha|\psi\rangle}{\langle \psi|\psi\rangle} = \langle \hat{P}_\alpha\rangle,
\end{equation*}
where we used the fact that the projector~\eqref{Seq: irrep projector} is Hermitian and squares to itself. 
Both $w_\alpha$ and $\langle \mathcal{P}\rangle$ were evaluated using 4\,000\,000 Monte Carlo samples again using~\eqref{Seq: local operators}. 
[We used different samples here, as NetKet does not offer a simple implementation of these symmetry operators, so it proved more expedient to sample $p(\vv\sigma)$ directly.]

Converged variational energies and all other observables are reported in Tables~\ref{tab: variational energies} and~\ref{tab: observables}, respectively.
%
%
%%%%%%%%%%%%%%%%%%%%%%%%%%%%%%%%%%%%%%%%%%%%%%%%%%%%%%%%%%%%%%

\section{Comparison of phase structure ans\"atze}
\label{Ssec: phase structures}

To demonstrate the advantage of our phase structure ansatz over other convolutional networks, we considered two alternative architectures:
\begin{enumerate}
    \item 24 convolutional filters spanning the entire lattice, followed by a ReLU activation layer and summation
    \item The architecture used for the amplitudes, except for the last layer, where the $\ln |x|$ activation is also replaced by ReLU.
\end{enumerate}
Amplitudes are encoded using the same ansatz as described in Appendix~\ref{Ssec: architecture}.
The performance of these architectures under the protocol described in Appendix~\ref{Ssec: protocol} is shown by the green and red curves in Fig.~\ref{Sfig: evolution}, respectively. 
In the first stage, neither of them approach the optimal variational energy found with the phasor sum ansatz;
subsequently, the amplitude network also fails to approach the ground state, even though it is capable of representing it closely, as found previously. 
We also point out that as the Monte Carlo sampling is restarted after changing the learning rate $\eta$, the estimates of the variational energy change substantially, leading to discontinuities in Fig.~\ref{Sfig: evolution}. This indicates that the amplitude structure had developed several unphysically strong peaks, which make subsequent Monte Carlo sampling unable to recover the correct wave function.

\begin{figure}
    \centering
    \includegraphics{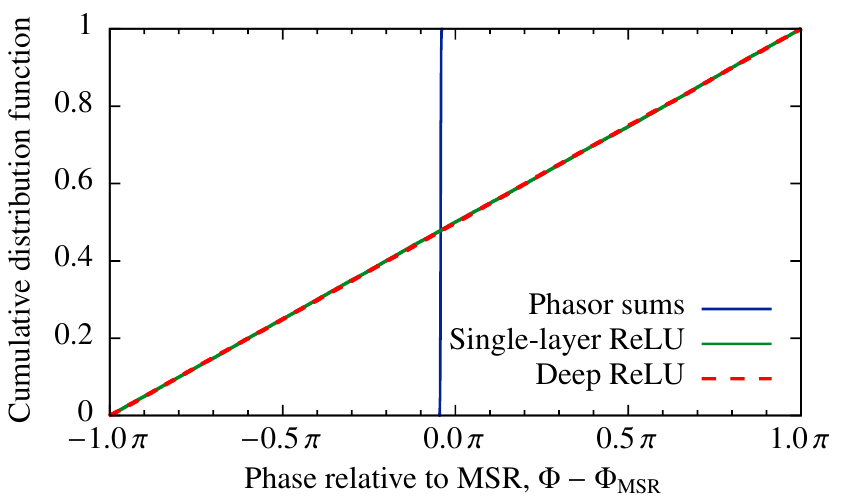}
    \caption{Cumulative distribution function of the Marshall-adjusted phases $\Phi-\Phi_\mathrm{MSR}$ learned by the three phase structure ans\"atze in the first stage of the optimisation in the unfrustrated case $J_2=0$. The phasor sum ansatz~\eqref{eq: phase machine} develops a very narrow distribution, that is, it reproduces the MSR to a good approximation; the other ans\"atze show an approximately uniform distribution, i.e., a complete failure to learn the sign rule.}
    \label{Sfig: MSR with bad ansatze}
\end{figure}

In the unfrustrated case, we also probe the phase structures learned by the various ans\"atze by comparing them directly to the exact Marshall sign rule. The distribution of the differences $\Phi-\Phi_\mathrm{MSR}$ is shown in Fig.~\ref{Sfig: MSR with bad ansatze}. While the architecture used in the main paper learns the MSR to a high accuracy (up to an irrelevant overall phase), the alternatives produce essentially random phases.
%
%
%%%%%%%%%%%%%%%%%%%%%%%%%%%%%%%%%%%%%%%%%%%%%%%%%%%%%%%%%%%%%%

\section{Average signs in exact diagonalisation}
\label{Ssec: average sign}

To estimate the average Marshall-adjusted sign of the true ground state of the $10\times10$ frustrated model, we obtained the exact ground states for $4\times4$ and $4\times6$ lattices using the Lanczos algorithm as implemented in SciPy~\cite{Virtanen2020SciPyPython}, and calculated the average sign
\begin{equation}
    \langle s \rangle_\psi = \left| \sum_{\vv\sigma} p_\psi(\vv\sigma) \frac{\langle\vv\sigma|\psi\rangle}{\big|\langle\vv\sigma|\psi\rangle\big|}\right| = \left| \frac{\sum_{\vv\sigma} \big|\langle\vv\sigma|\psi\rangle\big|\langle\vv\sigma|\psi\rangle}{\sum_{\vv\sigma} \big|\langle\vv\sigma|\psi\rangle\big|^2} \right|
\end{equation}
\begin{table}
    \centering
    \setlength{\tabcolsep}{1em}
    \iffalse
    \begin{tabular}{ccc}
         & $\langle s\rangle$ & $\langle s\rangle_\mathrm{MSR}$ \\\hline
        $4\times4$ & $3.53\times10^{-2}$ & 0.9745 \\
        $4\times6$ & $3.87\times10^{-3}$ & 0.9650 \\\hline\hline
        $10\times10$ & $\approx3\times10^{-12}$ & $\approx0.88$
    \end{tabular}
    \fi
    \begin{tabular}{cr@{${}\times{}$}lr@{.}l}
         & \multicolumn{2}{c}{$\langle s\rangle$} & \multicolumn{2}{c}{$\langle s\rangle_\mathrm{MSR}$} \\\hline
        $4\times4$ & 3.53&$10^{-2}$ & 0&9745 \\
        $4\times6$ & 3.87&$10^{-3}$ & 0&9650 \\\hline\hline
        $10\times10$ & $\approx3$&$10^{-12}$ & $\approx0$&88
    \end{tabular}
    \caption{Average sign $\langle s\rangle$ and Marshall-adjusted sign $\langle s\rangle_\mathrm{MSR}$ of the ground state of the $J_2/J_1=0.5$ square lattice HAFM for $4\times4$ and $4\times6$ lattices, calculated by exact diagonalisation. The $10\times10$ lattice is extrapolated from these, assuming that $\langle s\rangle$ decays exponentially in the number of spins~\cite{Troyer2005ComputationalSimulations}. }
    \label{Stab: average sign}
\end{table}
for both the original and the Marshall-adjusted ground states, $|\mathrm{GS}\rangle$ and $\prod_{i\in A} \sigma_i^z |\mathrm{GS}\rangle$. These average signs are given in Table~\ref{Stab: average sign}, together with an extrapolation to the $10\times10$ lattice, assuming an exponential decay of both $\langle s\rangle$~\cite{Troyer2005ComputationalSimulations}. The average sign of the original wave function decays extremely fast and becomes negligibly small for our lattice size. By contrast, the average Marshall-adjusted sign remains close, but clearly distinct from, 1. For a $10\times10$ lattice, the expected average sign is $\approx0.88$: Since this is the difference of the statistical weights of positive and negative amplitude states, we expect those to be about 94\% and 6\%, respectively.
%
%
%%%%%%%%%%%%%%%%%%%%%%%%%%%%%%%%%%%%%%%%%%%%%%%%%%%%%%%%%%%%%%

\section{Kernels of the ground state phase structures}
\label{Ssec: kernels}

\begin{figure*}
    \centering
    \includegraphics{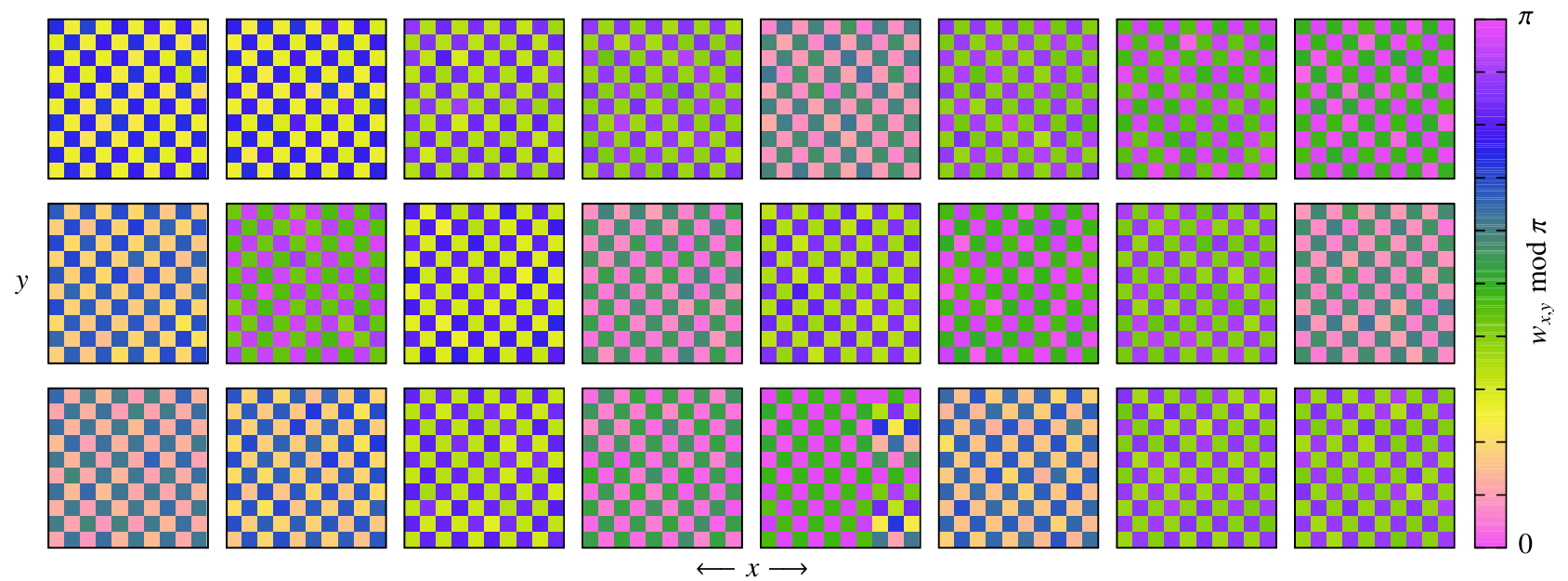}
    \includegraphics{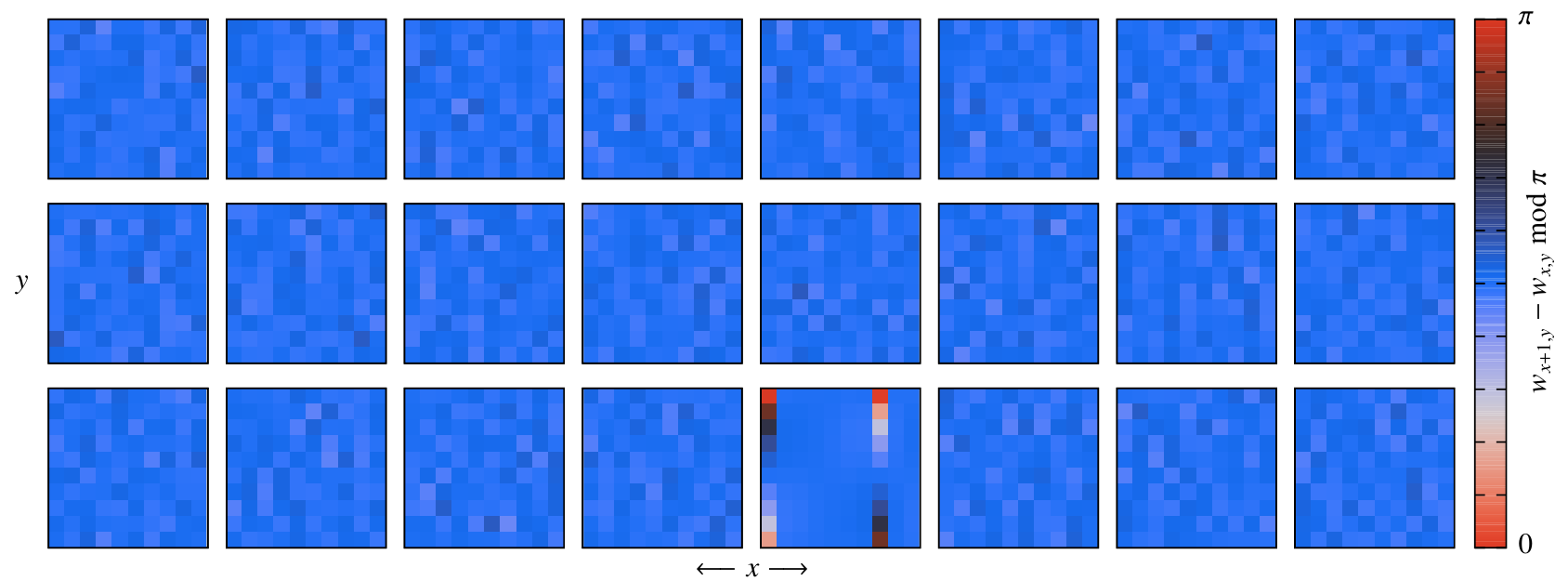}
    \includegraphics{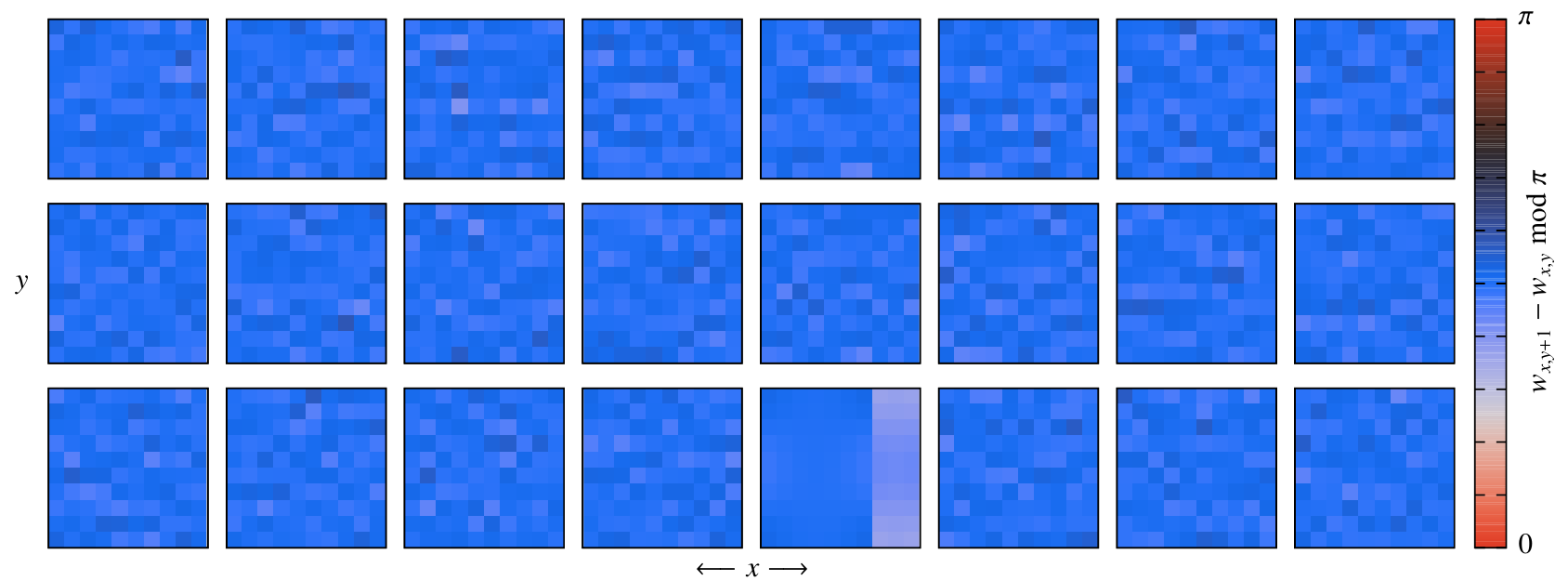}
    \caption{Weights $w_{n,\v r}$ of all convolutional kernels $n$ converged for $J_2=0$ (top three rows), and the differences $w_{n,\v r+\hat{\v x}}-w_{n,\v r}$ and $w_{n,\v r+\hat{\v y}}-w_{n,\v r}$ (middle and bottom three rows, respectively). 
    All kernels but one show a clear chequerboard pattern, which results in $\Delta\phi\approx\pi$ upon exchanging a neighbouring up and down spin, consistent with the Marshall sign rule. 
    In the only exception (fifth kernel of the third row), a ``topological fault'' spanning three columns appears, causing some $\Delta\phi$ to wind from 0 to $2\pi$: 
    these have little effect on the overall $\Delta\Phi$, and might persist due to a topologically invariant winding number.
    The kernel shown in Fig.~\ref{fig: kernels} is the first one in the second row.
    (Perceptionally uniform colour maps chosen following Ref.~\onlinecite{Kovesi2015GoodThem}).}
    \label{Sfig: MSR kernels}
\end{figure*}

\begin{figure*}
    \centering
    \includegraphics{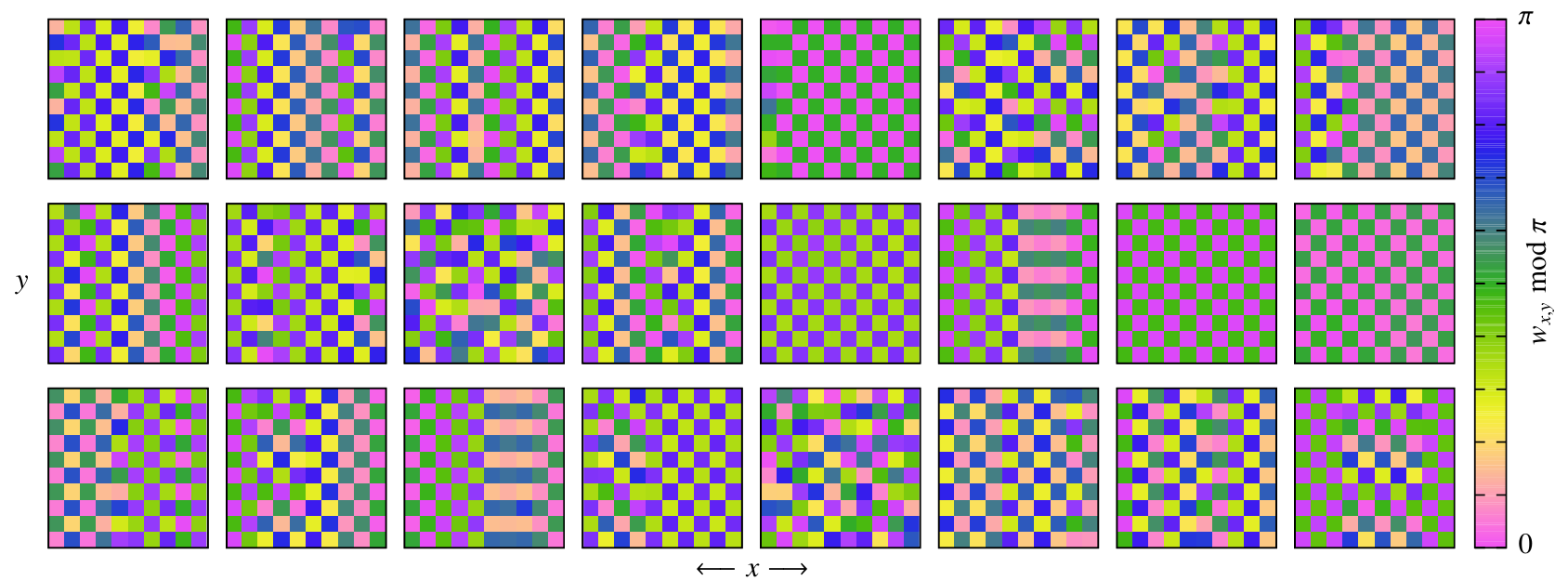}
    \includegraphics{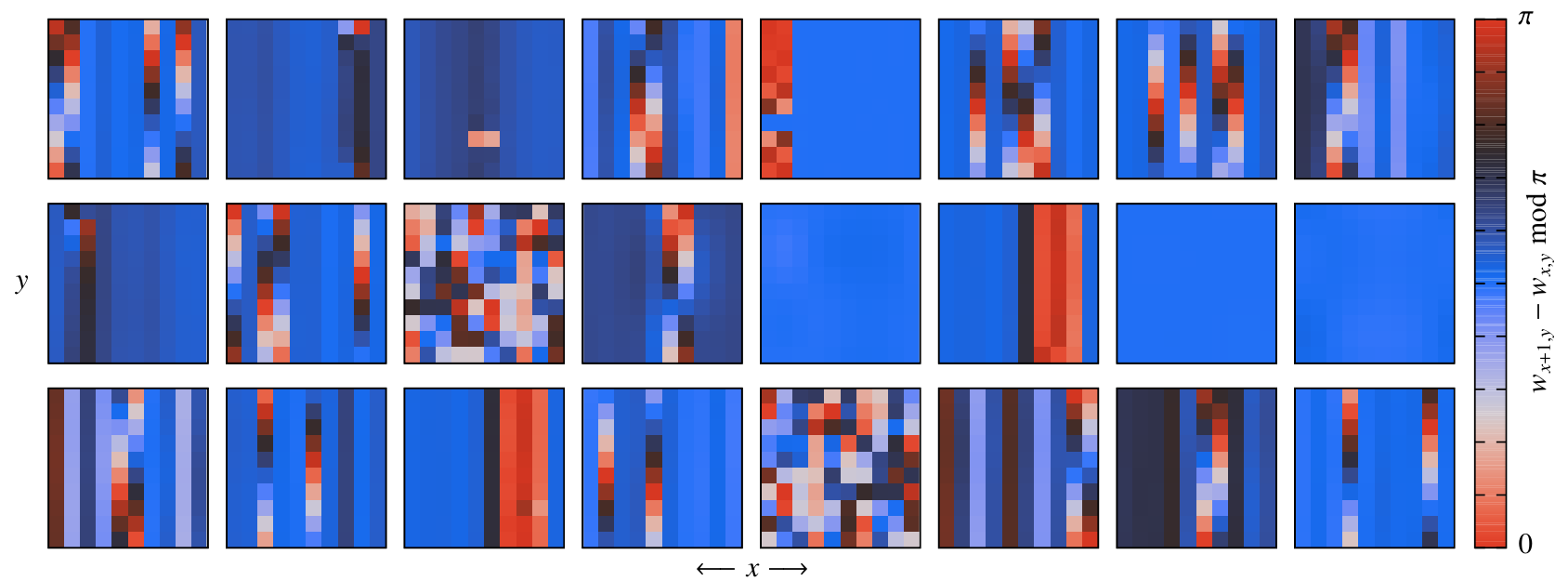}
    \includegraphics{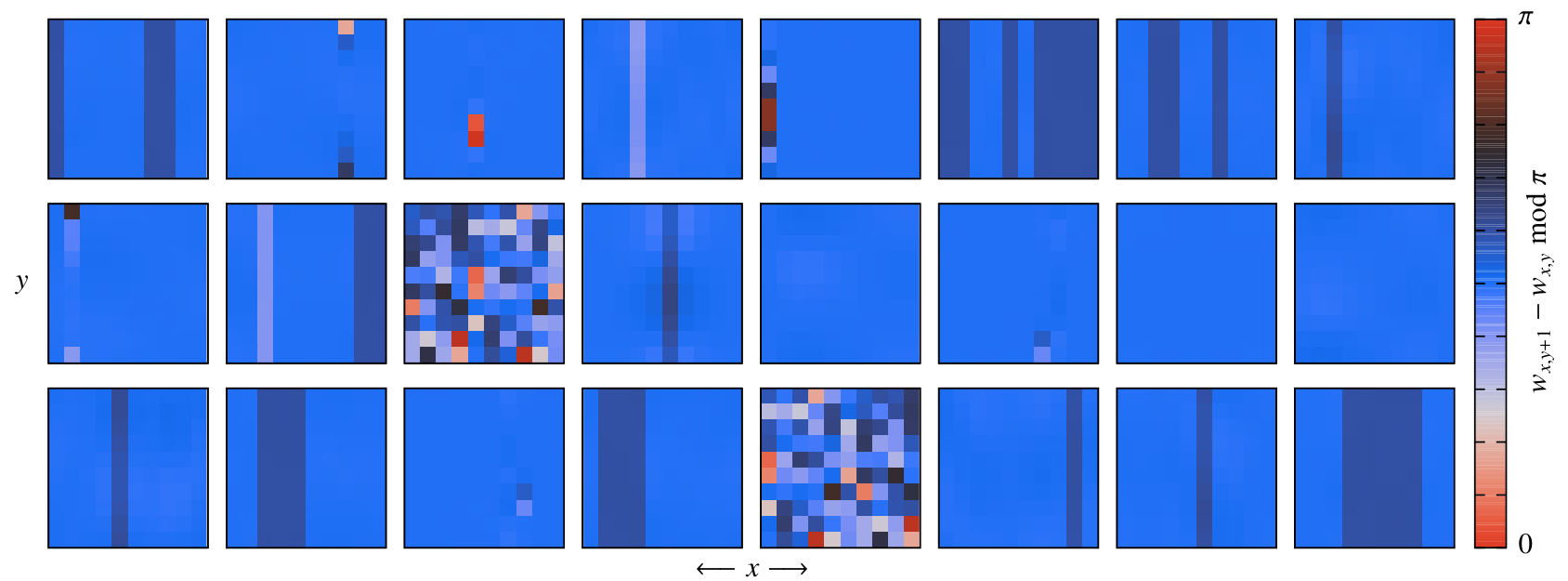}
    \caption{Weights $w_{n,\v r}$ of all convolutional kernels $n$ converged for $J_2/J_1=0.5$ (top three rows), and the differences $w_{n,\v r+\hat{\v x}}-w_{n,\v r}$ and $w_{n,\v r+\hat{\v y}}-w_{n,\v r}$ (middle and bottom three rows, respectively). 
    The kernels are altogether much less regular than in the unfrustrated case, several of them appearing completely scrambled, and many showing various winding patterns.
    Nonetheless, the chequerboard pattern corresponding to the MSR is still common.
    In addition, a stripy pattern with successive rows differing by $\pi/2$ appears, leading to $\Delta\phi\approx0$ upon exchanging nearest neighbour up and down spins horizontally but $\Delta\phi\approx\pi$ vertically, in an apparent breaking of rotational symmetry.
    The kernel shown in Fig.~\ref{fig: kernels} is the sixth one in the second row.
    (Perceptionally uniform colour maps chosen following Ref.~\onlinecite{Kovesi2015GoodThem}.)}
    \label{Sfig: frustrated kernels}
\end{figure*}

As discussed in Sec.~\ref{sec: discussion}, the change in the phase ansatz
%\begin{equation}
%    \Phi(\vv\sigma) = \arg\bigg[\sum_{n,\v r} \exp\Big( i b_n +  \sum_{\v r'} iw_{n,\v r-\v r'}\sigma^z_{\v r'} \Big)\bigg]
%    \label{Seq: phase machine}
%\end{equation}
\eqref{eq: phase machine}
upon exchanging an up and a down spin separated by $\v R$ is a kind of average of the change $\Delta\phi = 2(w_{n,\v r+\v R} - w_{n,\v r})$ in each elementary phase that enters it. %~\eqref{Seq: phase machine}. 
(The factor of 2 is due to representing up and down spins as $\pm1$ rather than $\pm1/2$.)
Since the energy $+J\sigma_i^+\sigma_j^-$ of an antiferromagnetic interaction is optimised if $\Delta\Phi=\pi$ upon exchanging spins $i$ and $j$, we expect the corresponding set of $\Delta\phi$ to have a distribution centred upon $\pi$, especially in the unfrustrated case, where all interactions in the Hamiltonian can be optimised simultaneously.

In particular, one can represent the Marshall sign rule corresponding to the unfrustrated limit $J_2=0$ exactly by requiring that all $\Delta\phi\equiv\pi\pmod{2\pi}$, i.e., $w_{\v r+\v R}-w_{\v r}\equiv\pi/2\pmod\pi$ for all nearest neighbour pairs, or $\v R = \hat{\v x}, \hat{\v y}$. 
This is readily achieved by a chequerboard pattern of weights $w$ and $w+\pi/2$, provided both sides of the lattice are of even length. 
Indeed, the phase of each elementary phasor in this setup is
\begin{align}
    \phi_{\v r} &= \sum_{\v r'} w_{n,\v r-\v r'}\sigma^z_{\v r'} = w\sum_{\v r'}\sigma^z_{\v r'} + \frac\pi2\sum_{\v r'\in A(B)}\sigma^z_{\v r'} \nonumber\\
    &= -\frac{\pi N}4 + \pi \sum_{\v r'\in B(A)} \frac{1-\sigma^z_{\v r'}}2 \nonumber\\
    &\equiv \frac{\pi N}4 + \pi \sum_{\v r'\in A} \frac{1-\sigma^z_{\v r'}}2 \pmod{2\pi},
\end{align}
where $N$ is the number of lattice sites, $A$ and $B$ are the two chequerboard sublattices of the lattice, and we repeatedly use the fact that $\sum\sigma^z = 0$. In the last line, we also note that the total number of down spins (which is measured by the sums for each sublattice) is $N/2$, which is even; therefore, the parity of the sum for the two sublattices is the same. 
That is, all terms in~\eqref{eq: phase machine} have the same phase, which is also consistent with the MSR~\eqref{eq: Marshall sign rule} up to an overall phase $\pi N/4$. It follows that $\Phi$ equals this phase, and thus recovers the MSR.

Beyond the nearest neighbour case, there is a Marshall sign rule corresponding to the other unfrustrated limit, $J_1=0$, arising from the requirement that $\Delta\Phi=\pi$ upon exchanging next-nearest neighbour up and down spins. 
At this point, the two chequerboard sublattices decouple, each being effectively a square lattice with diagonal axes.
Using the above construction, appropriate kernels $w_{\v r}$ can be found for both of them; however, the offset between the two sublattices is arbitrary, as it only contributes to an unimportant overall phase. 
Upon reintroducing a weak nearest-neighbour coupling, however, the model develops a stripy antiferromagnetic order~\cite{Hu2013DirectAntiferromagnetism,Gong2014PlaquetteModel}.
Consistently, the sublattices in the convolutional weights $w$ are expected to lock so as to form rows or columns with equal $w$, shifted from one another by $\pi/2\pmod\pi$.

We now consider the convolutional kernels generated by the variational Monte Carlo protocol both in the unfrustrated limit and at $J_2/J_1=0.5$. The weights $w_{n,\v r}$ are plotted for each kernel in Figs.~\ref{Sfig: MSR kernels} and~\ref{Sfig: frustrated kernels}, respectively. In both cases, we also plot the differences between horizontally and vertically nearest-neighbour kernel entries.

At $J_2=0$, the chequerboard pattern derived for the exact MSR can be seen in almost all kernels to a very good approximation; consistently, $\Delta\phi\approx\pi$ upon exchanging nearest neighbour spins, both horizontally and vertically.
This suggests that this representation of the MSR is especially stable; presumably, it sits at the bottom of a wide basin of the learning landscape, making it easy to find for optimisation algorithms~\cite{Baldassi2020ShapingMinima,Goodfellow2016DeepLearning}. 
The only surprising feature in Fig.~\ref{Sfig: MSR kernels} is a kernel that develops a three column wide ``topological fault,'' in which the Marshall-adjusted convolutional weights wind around the vertical direction. 
This results in a number of $\Delta\phi$ far from the desired $\pi$.
The large number of kernels, however, allows these to be corrected through slight deviations of the other kernels from the exact MSR.
Furthermore, the additional kernels probably play a key role in eliminating such detrimental structures:
Unwinding a ``topological fault'' requires large-scale changes in the individual phasors, which substantially increase the variational energy, unless corrected for by the other kernels.
Indeed, our attempts to use a single convolutional kernel in~\eqref{eq: phase machine} were plagued by robust ``topological faults'' spanning the entire kernel, leading to convergence far above the ground state energy.

The kernels obtained in the frustrated case are substantially more complex, with many ``topological faults'' and some kernels that show no discernible pattern. 
Many kernels, however, retain the chequerboard pattern consistent with the MSR, and the $\Delta\phi$ upon exchanging nearest neighbour spins vertically are clearly dominated by values close to $\pi$. 
Both of these are consistent with the fact that the kernels represent the MSR rather than the true frustrated sign structure. 
Nevertheless, we observe several columns of the stripy pattern consistent with the MSR of the $J_1=0$ limit, as discussed above. 
These result in $\Delta\phi\approx0$ for horizontal nearest neighbour exchanges, leading to a much more varied pattern in these differences.
It is surprising that this more diverse distribution has no apparent effect on the overall sign structure learned by the network.

Furthermore, the striking difference between $\Delta\phi$ along the horizontal and the vertical directions is not warranted either by any fundamental property of the ansatz, or by the final converged wave function that obeys all point group symmetries to a high accuracy (see Table~\ref{tab: observables}). Nevertheless, it might hint at spontaneous point group symmetry breaking at higher variational energies that ultimately leads to learning an incorrect sign structure. 
A more detailed analysis of the learning dynamics is necessary to better understand and overcome any such problems.
%
%
%%%%%%%%%%%%%%%%%%%%%%%%%%%%%%%%%%%%%%%%%%%%%%%%%%%%%%%%%%%%%%

\bibliography{references}

\begin{thebibliography}{49}%
\makeatletter
\providecommand \@ifxundefined [1]{%
 \@ifx{#1\undefined}
}%
\providecommand \@ifnum [1]{%
 \ifnum #1\expandafter \@firstoftwo
 \else \expandafter \@secondoftwo
 \fi
}%
\providecommand \@ifx [1]{%
 \ifx #1\expandafter \@firstoftwo
 \else \expandafter \@secondoftwo
 \fi
}%
\providecommand \natexlab [1]{#1}%
\providecommand \enquote  [1]{``#1''}%
\providecommand \bibnamefont  [1]{#1}%
\providecommand \bibfnamefont [1]{#1}%
\providecommand \citenamefont [1]{#1}%
\providecommand \href@noop [0]{\@secondoftwo}%
\providecommand \href [0]{\begingroup \@sanitize@url \@href}%
\providecommand \@href[1]{\@@startlink{#1}\@@href}%
\providecommand \@@href[1]{\endgroup#1\@@endlink}%
\providecommand \@sanitize@url [0]{\catcode `\\12\catcode `\$12\catcode
  `\&12\catcode `\#12\catcode `\^12\catcode `\_12\catcode `\%12\relax}%
\providecommand \@@startlink[1]{}%
\providecommand \@@endlink[0]{}%
\providecommand \url  [0]{\begingroup\@sanitize@url \@url }%
\providecommand \@url [1]{\endgroup\@href {#1}{\urlprefix }}%
\providecommand \urlprefix  [0]{URL }%
\providecommand \Eprint [0]{\href }%
\providecommand \doibase [0]{http://dx.doi.org/}%
\providecommand \selectlanguage [0]{\@gobble}%
\providecommand \bibinfo  [0]{\@secondoftwo}%
\providecommand \bibfield  [0]{\@secondoftwo}%
\providecommand \translation [1]{[#1]}%
\providecommand \BibitemOpen [0]{}%
\providecommand \bibitemStop [0]{}%
\providecommand \bibitemNoStop [0]{.\EOS\space}%
\providecommand \EOS [0]{\spacefactor3000\relax}%
\providecommand \BibitemShut  [1]{\csname bibitem#1\endcsname}%
\let\auto@bib@innerbib\@empty
%</preamble>
\bibitem [{\citenamefont {Carleo}\ \emph
  {et~al.}(2019{\natexlab{a}})\citenamefont {Carleo}, \citenamefont {Cirac},
  \citenamefont {Cranmer}, \citenamefont {Daudet}, \citenamefont {Schuld},
  \citenamefont {Tishby}, \citenamefont {Vogt-Maranto},\ and\ \citenamefont
  {Zdeborov{\'{a}}}}]{Carleo2019MachineSciences}%
  \BibitemOpen
  \bibfield  {author} {\bibinfo {author} {\bibfnamefont {G.}~\bibnamefont
  {Carleo}}, \bibinfo {author} {\bibfnamefont {I.}~\bibnamefont {Cirac}},
  \bibinfo {author} {\bibfnamefont {K.}~\bibnamefont {Cranmer}}, \bibinfo
  {author} {\bibfnamefont {L.}~\bibnamefont {Daudet}}, \bibinfo {author}
  {\bibfnamefont {M.}~\bibnamefont {Schuld}}, \bibinfo {author} {\bibfnamefont
  {N.}~\bibnamefont {Tishby}}, \bibinfo {author} {\bibfnamefont
  {L.}~\bibnamefont {Vogt-Maranto}}, \ and\ \bibinfo {author} {\bibfnamefont
  {L.}~\bibnamefont {Zdeborov{\'{a}}}},\ }\href {\doibase
  10.1103/RevModPhys.91.045002} {\bibfield  {journal} {\bibinfo  {journal}
  {Rev. Mod. Phys.}\ }\textbf {\bibinfo {volume} {91}},\ \bibinfo {pages}
  {045002} (\bibinfo {year} {2019}{\natexlab{a}})}\BibitemShut {NoStop}%
\bibitem [{\citenamefont {Carleo}\ and\ \citenamefont
  {Troyer}(2017)}]{Carleo2017SolvingNetworks}%
  \BibitemOpen
  \bibfield  {author} {\bibinfo {author} {\bibfnamefont {G.}~\bibnamefont
  {Carleo}}\ and\ \bibinfo {author} {\bibfnamefont {M.}~\bibnamefont
  {Troyer}},\ }\href {\doibase 10.1126/science.aag2302} {\bibfield  {journal}
  {\bibinfo  {journal} {Science}\ }\textbf {\bibinfo {volume} {355}},\ \bibinfo
  {pages} {602} (\bibinfo {year} {2017})}\BibitemShut {NoStop}%
\bibitem [{\citenamefont {Melko}\ \emph {et~al.}(2019)\citenamefont {Melko},
  \citenamefont {Carleo}, \citenamefont {Carrasquilla},\ and\ \citenamefont
  {Cirac}}]{Melko2019RestrictedPhysics}%
  \BibitemOpen
  \bibfield  {author} {\bibinfo {author} {\bibfnamefont {R.~G.}\ \bibnamefont
  {Melko}}, \bibinfo {author} {\bibfnamefont {G.}~\bibnamefont {Carleo}},
  \bibinfo {author} {\bibfnamefont {J.}~\bibnamefont {Carrasquilla}}, \ and\
  \bibinfo {author} {\bibfnamefont {J.~I.}\ \bibnamefont {Cirac}},\ }\href
  {\doibase 10.1038/s41567-019-0545-1} {\bibfield  {journal} {\bibinfo
  {journal} {Nat. Phys.}\ }\textbf {\bibinfo {volume} {15}},\ \bibinfo {pages}
  {887} (\bibinfo {year} {2019})}\BibitemShut {NoStop}%
\bibitem [{\citenamefont {Choo}\ \emph {et~al.}(2019)\citenamefont {Choo},
  \citenamefont {Neupert},\ and\ \citenamefont {Carleo}}]{Choo2019StudyStates}%
  \BibitemOpen
  \bibfield  {author} {\bibinfo {author} {\bibfnamefont {K.}~\bibnamefont
  {Choo}}, \bibinfo {author} {\bibfnamefont {T.}~\bibnamefont {Neupert}}, \
  and\ \bibinfo {author} {\bibfnamefont {G.}~\bibnamefont {Carleo}},\ }\href
  {\doibase 10.1103/PhysRevB.100.125124} {\bibfield  {journal} {\bibinfo
  {journal} {Phys. Rev. B}\ }\textbf {\bibinfo {volume} {100}},\ \bibinfo
  {pages} {125124} (\bibinfo {year} {2019})}\BibitemShut {NoStop}%
\bibitem [{\citenamefont {Choo}\ \emph {et~al.}(2020)\citenamefont {Choo},
  \citenamefont {Mezzacapo},\ and\ \citenamefont
  {Carleo}}]{Choo2020FermionicStructure}%
  \BibitemOpen
  \bibfield  {author} {\bibinfo {author} {\bibfnamefont {K.}~\bibnamefont
  {Choo}}, \bibinfo {author} {\bibfnamefont {A.}~\bibnamefont {Mezzacapo}}, \
  and\ \bibinfo {author} {\bibfnamefont {G.}~\bibnamefont {Carleo}},\ }\href
  {\doibase 10.1038/s41467-020-15724-9} {\bibfield  {journal} {\bibinfo
  {journal} {Nat. Comm.}\ }\textbf {\bibinfo {volume} {11}},\ \bibinfo {pages}
  {2368} (\bibinfo {year} {2020})}\BibitemShut {NoStop}%
\bibitem [{\citenamefont {Sharir}\ \emph {et~al.}(2020)\citenamefont {Sharir},
  \citenamefont {Levine}, \citenamefont {Wies}, \citenamefont {Carleo},\ and\
  \citenamefont {Shashua}}]{Sharir2020DeepSystems}%
  \BibitemOpen
  \bibfield  {author} {\bibinfo {author} {\bibfnamefont {O.}~\bibnamefont
  {Sharir}}, \bibinfo {author} {\bibfnamefont {Y.}~\bibnamefont {Levine}},
  \bibinfo {author} {\bibfnamefont {N.}~\bibnamefont {Wies}}, \bibinfo {author}
  {\bibfnamefont {G.}~\bibnamefont {Carleo}}, \ and\ \bibinfo {author}
  {\bibfnamefont {A.}~\bibnamefont {Shashua}},\ }\href {\doibase
  10.1103/PhysRevLett.124.020503} {\bibfield  {journal} {\bibinfo  {journal}
  {Phys. Rev. Lett.}\ }\textbf {\bibinfo {volume} {124}},\ \bibinfo {pages}
  {020503} (\bibinfo {year} {2020})}\BibitemShut {NoStop}%
\bibitem [{\citenamefont {Yang}\ \emph {et~al.}(2020)\citenamefont {Yang},
  \citenamefont {Leng}, \citenamefont {Yu}, \citenamefont {Patel},
  \citenamefont {Hu},\ and\ \citenamefont {Pu}}]{Yang2020DeepPhysics}%
  \BibitemOpen
  \bibfield  {author} {\bibinfo {author} {\bibfnamefont {L.}~\bibnamefont
  {Yang}}, \bibinfo {author} {\bibfnamefont {Z.}~\bibnamefont {Leng}}, \bibinfo
  {author} {\bibfnamefont {G.}~\bibnamefont {Yu}}, \bibinfo {author}
  {\bibfnamefont {A.}~\bibnamefont {Patel}}, \bibinfo {author} {\bibfnamefont
  {W.-J.}\ \bibnamefont {Hu}}, \ and\ \bibinfo {author} {\bibfnamefont
  {H.}~\bibnamefont {Pu}},\ }\href {\doibase 10.1103/physrevresearch.2.012039}
  {\bibfield  {journal} {\bibinfo  {journal} {Phys. Rev. Research}\ }\textbf
  {\bibinfo {volume} {2}},\ \bibinfo {pages} {012039} (\bibinfo {year}
  {2020})}\BibitemShut {NoStop}%
\bibitem [{\citenamefont {Le~Roux}\ and\ \citenamefont
  {Bengio}(2008)}]{LeRoux2008RepresentationalNetworks}%
  \BibitemOpen
  \bibfield  {author} {\bibinfo {author} {\bibfnamefont {N.}~\bibnamefont
  {Le~Roux}}\ and\ \bibinfo {author} {\bibfnamefont {Y.}~\bibnamefont
  {Bengio}},\ }\href {\doibase 10.1162/neco.2008.04-07-510} {\bibfield
  {journal} {\bibinfo  {journal} {Neural Comput.}\ }\textbf {\bibinfo {volume}
  {20}},\ \bibinfo {pages} {1631} (\bibinfo {year} {2008})}\BibitemShut
  {NoStop}%
\bibitem [{\citenamefont {Cybenko}(1989)}]{Cybenko1989ApproximationFunction}%
  \BibitemOpen
  \bibfield  {author} {\bibinfo {author} {\bibfnamefont {G.}~\bibnamefont
  {Cybenko}},\ }\href {\doibase 10.1007/BF02551274} {\bibfield  {journal}
  {\bibinfo  {journal} {Math. Control Signals Syst.}\ }\textbf {\bibinfo
  {volume} {2}},\ \bibinfo {pages} {303} (\bibinfo {year} {1989})}\BibitemShut
  {NoStop}%
\bibitem [{\citenamefont {Li}\ and\ \citenamefont
  {Yao}(2019)}]{Li2019Sign-Problem-FreeApplications}%
  \BibitemOpen
  \bibfield  {author} {\bibinfo {author} {\bibfnamefont {Z.-X.}\ \bibnamefont
  {Li}}\ and\ \bibinfo {author} {\bibfnamefont {H.}~\bibnamefont {Yao}},\
  }\href {\doibase 10.1146/annurev-conmatphys-033117-054307} {\bibfield
  {journal} {\bibinfo  {journal} {Annu. Rev. Condens. Matter Phys.}\ }\textbf
  {\bibinfo {volume} {10}},\ \bibinfo {pages} {337} (\bibinfo {year}
  {2019})}\BibitemShut {NoStop}%
\bibitem [{\citenamefont {Wolf}\ \emph {et~al.}(2008)\citenamefont {Wolf},
  \citenamefont {Verstraete}, \citenamefont {Hastings},\ and\ \citenamefont
  {Cirac}}]{Wolf2008AreaCorrelations}%
  \BibitemOpen
  \bibfield  {author} {\bibinfo {author} {\bibfnamefont {M.~M.}\ \bibnamefont
  {Wolf}}, \bibinfo {author} {\bibfnamefont {F.}~\bibnamefont {Verstraete}},
  \bibinfo {author} {\bibfnamefont {M.~B.}\ \bibnamefont {Hastings}}, \ and\
  \bibinfo {author} {\bibfnamefont {J.~I.}\ \bibnamefont {Cirac}},\ }\href
  {\doibase 10.1103/PhysRevLett.100.070502} {\bibfield  {journal} {\bibinfo
  {journal} {Phys. Rev. Lett.}\ }\textbf {\bibinfo {volume} {100}},\ \bibinfo
  {pages} {070502} (\bibinfo {year} {2008})}\BibitemShut {NoStop}%
\bibitem [{\citenamefont {Riera}\ and\ \citenamefont
  {Latorre}(2006)}]{Riera2006AreaNetworks}%
  \BibitemOpen
  \bibfield  {author} {\bibinfo {author} {\bibfnamefont {A.}~\bibnamefont
  {Riera}}\ and\ \bibinfo {author} {\bibfnamefont {J.~I.}\ \bibnamefont
  {Latorre}},\ }\href {\doibase 10.1103/PhysRevA.74.052326} {\bibfield
  {journal} {\bibinfo  {journal} {Phys. Rev. A}\ }\textbf {\bibinfo {volume}
  {74}},\ \bibinfo {pages} {052326} (\bibinfo {year} {2006})}\BibitemShut
  {NoStop}%
\bibitem [{\citenamefont {Or{\'{u}}s}(2019)}]{Orus2019TensorSystems}%
  \BibitemOpen
  \bibfield  {author} {\bibinfo {author} {\bibfnamefont {R.}~\bibnamefont
  {Or{\'{u}}s}},\ }\href {\doibase 10.1038/s42254-019-0086-7} {\bibfield
  {journal} {\bibinfo  {journal} {Nat. Rev. Phys.}\ }\textbf {\bibinfo {volume}
  {1}},\ \bibinfo {pages} {538} (\bibinfo {year} {2019})}\BibitemShut {NoStop}%
\bibitem [{\citenamefont {Levine}\ \emph {et~al.}(2019)\citenamefont {Levine},
  \citenamefont {Sharir}, \citenamefont {Cohen},\ and\ \citenamefont
  {Shashua}}]{Levine2019QuantumArchitectures}%
  \BibitemOpen
  \bibfield  {author} {\bibinfo {author} {\bibfnamefont {Y.}~\bibnamefont
  {Levine}}, \bibinfo {author} {\bibfnamefont {O.}~\bibnamefont {Sharir}},
  \bibinfo {author} {\bibfnamefont {N.}~\bibnamefont {Cohen}}, \ and\ \bibinfo
  {author} {\bibfnamefont {A.}~\bibnamefont {Shashua}},\ }\href {\doibase
  10.1103/PhysRevLett.122.065301} {\bibfield  {journal} {\bibinfo  {journal}
  {Phys. Rev. Lett.}\ }\textbf {\bibinfo {volume} {122}},\ \bibinfo {pages}
  {065301} (\bibinfo {year} {2019})}\BibitemShut {NoStop}%
\bibitem [{\citenamefont {Kaubruegger}\ \emph {et~al.}(2018)\citenamefont
  {Kaubruegger}, \citenamefont {Pastori},\ and\ \citenamefont
  {Budich}}]{Kaubruegger2018ChiralNetworks}%
  \BibitemOpen
  \bibfield  {author} {\bibinfo {author} {\bibfnamefont {R.}~\bibnamefont
  {Kaubruegger}}, \bibinfo {author} {\bibfnamefont {L.}~\bibnamefont
  {Pastori}}, \ and\ \bibinfo {author} {\bibfnamefont {J.~C.}\ \bibnamefont
  {Budich}},\ }\href {\doibase 10.1103/PhysRevB.97.195136} {\bibfield
  {journal} {\bibinfo  {journal} {Phys. Rev. B}\ }\textbf {\bibinfo {volume}
  {97}},\ \bibinfo {pages} {195136} (\bibinfo {year} {2018})}\BibitemShut
  {NoStop}%
\bibitem [{\citenamefont {Choo}\ \emph {et~al.}(2018)\citenamefont {Choo},
  \citenamefont {Carleo}, \citenamefont {Regnault},\ and\ \citenamefont
  {Neupert}}]{Choo2018SymmetriesStates}%
  \BibitemOpen
  \bibfield  {author} {\bibinfo {author} {\bibfnamefont {K.}~\bibnamefont
  {Choo}}, \bibinfo {author} {\bibfnamefont {G.}~\bibnamefont {Carleo}},
  \bibinfo {author} {\bibfnamefont {N.}~\bibnamefont {Regnault}}, \ and\
  \bibinfo {author} {\bibfnamefont {T.}~\bibnamefont {Neupert}},\ }\href
  {\doibase 10.1103/PhysRevLett.121.167204} {\bibfield  {journal} {\bibinfo
  {journal} {Phys. Rev. Lett.}\ }\textbf {\bibinfo {volume} {121}},\ \bibinfo
  {pages} {167204} (\bibinfo {year} {2018})}\BibitemShut {NoStop}%
\bibitem [{\citenamefont {Cai}\ and\ \citenamefont
  {Liu}(2018)}]{Cai2018ApproximatingNetworks}%
  \BibitemOpen
  \bibfield  {author} {\bibinfo {author} {\bibfnamefont {Z.}~\bibnamefont
  {Cai}}\ and\ \bibinfo {author} {\bibfnamefont {J.}~\bibnamefont {Liu}},\
  }\href {\doibase 10.1103/PhysRevB.97.035116} {\bibfield  {journal} {\bibinfo
  {journal} {Phys. Rev. B}\ }\textbf {\bibinfo {volume} {97}},\ \bibinfo
  {pages} {035116} (\bibinfo {year} {2018})}\BibitemShut {NoStop}%
\bibitem [{\citenamefont {Hendry}\ and\ \citenamefont
  {Feiguin}(2019)}]{Hendry2019MachineSystems}%
  \BibitemOpen
  \bibfield  {author} {\bibinfo {author} {\bibfnamefont {D.}~\bibnamefont
  {Hendry}}\ and\ \bibinfo {author} {\bibfnamefont {A.~E.}\ \bibnamefont
  {Feiguin}},\ }\href {\doibase 10.1103/PhysRevB.100.245123} {\bibfield
  {journal} {\bibinfo  {journal} {Phys. Rev. B}\ }\textbf {\bibinfo {volume}
  {100}},\ \bibinfo {pages} {245123} (\bibinfo {year} {2019})}\BibitemShut
  {NoStop}%
\bibitem [{\citenamefont {Torlai}\ \emph {et~al.}(2019)\citenamefont {Torlai},
  \citenamefont {Carrasquilla}, \citenamefont {Fishman}, \citenamefont
  {Melko},\ and\ \citenamefont
  {Fisher}}]{Torlai2019WavefunctionDifferentiation}%
  \BibitemOpen
  \bibfield  {author} {\bibinfo {author} {\bibfnamefont {G.}~\bibnamefont
  {Torlai}}, \bibinfo {author} {\bibfnamefont {J.}~\bibnamefont
  {Carrasquilla}}, \bibinfo {author} {\bibfnamefont {M.~T.}\ \bibnamefont
  {Fishman}}, \bibinfo {author} {\bibfnamefont {R.~G.}\ \bibnamefont {Melko}},
  \ and\ \bibinfo {author} {\bibfnamefont {M.~P.~A.}\ \bibnamefont {Fisher}},\
  }\href {http://arxiv.org/abs/1906.04654} {\bibfield  {journal} {\bibinfo
  {journal} {arXiv:1906.04654}\ } (\bibinfo {year} {2019})}\BibitemShut
  {NoStop}%
\bibitem [{\citenamefont {Westerhout}\ \emph {et~al.}(2020)\citenamefont
  {Westerhout}, \citenamefont {Astrakhantsev}, \citenamefont {Tikhonov},
  \citenamefont {Katsnelson},\ and\ \citenamefont
  {Bagrov}}]{Westerhout2020GeneralizationStates}%
  \BibitemOpen
  \bibfield  {author} {\bibinfo {author} {\bibfnamefont {T.}~\bibnamefont
  {Westerhout}}, \bibinfo {author} {\bibfnamefont {N.}~\bibnamefont
  {Astrakhantsev}}, \bibinfo {author} {\bibfnamefont {K.~S.}\ \bibnamefont
  {Tikhonov}}, \bibinfo {author} {\bibfnamefont {M.~I.}\ \bibnamefont
  {Katsnelson}}, \ and\ \bibinfo {author} {\bibfnamefont {A.~A.}\ \bibnamefont
  {Bagrov}},\ }\href {\doibase 10.1038/s41467-020-15402-w} {\bibfield
  {journal} {\bibinfo  {journal} {Nat. Comm.}\ }\textbf {\bibinfo {volume}
  {11}},\ \bibinfo {pages} {1593} (\bibinfo {year} {2020})}\BibitemShut
  {NoStop}%
\bibitem [{\citenamefont {Nomura}(2020)}]{Nomura2020MachineCalculations}%
  \BibitemOpen
  \bibfield  {author} {\bibinfo {author} {\bibfnamefont {Y.}~\bibnamefont
  {Nomura}},\ }\href {\doibase 10.7566/JPSJ.89.054706} {\bibfield  {journal}
  {\bibinfo  {journal} {J. Phys. Soc. Jpn.}\ }\textbf {\bibinfo {volume}
  {89}},\ \bibinfo {pages} {054706} (\bibinfo {year} {2020})}\BibitemShut
  {NoStop}%
\bibitem [{\citenamefont {d'Ascoli}\ \emph {et~al.}(2019)\citenamefont
  {d'Ascoli}, \citenamefont {Sagun}, \citenamefont {Bruna},\ and\ \citenamefont
  {Biroli}}]{dAscoli2019FindingBias}%
  \BibitemOpen
  \bibfield  {author} {\bibinfo {author} {\bibfnamefont {S.}~\bibnamefont
  {d'Ascoli}}, \bibinfo {author} {\bibfnamefont {L.}~\bibnamefont {Sagun}},
  \bibinfo {author} {\bibfnamefont {J.}~\bibnamefont {Bruna}}, \ and\ \bibinfo
  {author} {\bibfnamefont {G.}~\bibnamefont {Biroli}},\ }\href
  {http://arxiv.org/abs/1906.06766} {\bibfield  {journal} {\bibinfo  {journal}
  {arXiv:1906.06766}\ } (\bibinfo {year} {2019})}\BibitemShut {NoStop}%
\bibitem [{\citenamefont {Marshall}\ and\ \citenamefont
  {Peierls}(1955)}]{Marshall1955Antiferromagnetism}%
  \BibitemOpen
  \bibfield  {author} {\bibinfo {author} {\bibfnamefont {W.}~\bibnamefont
  {Marshall}}\ and\ \bibinfo {author} {\bibfnamefont {R.~E.}\ \bibnamefont
  {Peierls}},\ }\href {\doibase 10.1098/rspa.1955.0200} {\bibfield  {journal}
  {\bibinfo  {journal} {Proc. R. Soc. London A}\ }\textbf {\bibinfo {volume}
  {232}},\ \bibinfo {pages} {48} (\bibinfo {year} {1955})}\BibitemShut
  {NoStop}%
\bibitem [{\citenamefont {Liang}\ \emph {et~al.}(2018)\citenamefont {Liang},
  \citenamefont {Liu}, \citenamefont {Lin}, \citenamefont {Guo}, \citenamefont
  {Zhang},\ and\ \citenamefont {He}}]{Liang2018SolvingNetworks}%
  \BibitemOpen
  \bibfield  {author} {\bibinfo {author} {\bibfnamefont {X.}~\bibnamefont
  {Liang}}, \bibinfo {author} {\bibfnamefont {W.-Y.}\ \bibnamefont {Liu}},
  \bibinfo {author} {\bibfnamefont {P.-Z.}\ \bibnamefont {Lin}}, \bibinfo
  {author} {\bibfnamefont {G.-C.}\ \bibnamefont {Guo}}, \bibinfo {author}
  {\bibfnamefont {Y.-S.}\ \bibnamefont {Zhang}}, \ and\ \bibinfo {author}
  {\bibfnamefont {L.}~\bibnamefont {He}},\ }\href {\doibase
  10.1103/PhysRevB.98.104426} {\bibfield  {journal} {\bibinfo  {journal} {Phys.
  Rev. B}\ }\textbf {\bibinfo {volume} {98}},\ \bibinfo {pages} {104426}
  (\bibinfo {year} {2018})}\BibitemShut {NoStop}%
\bibitem [{\citenamefont {Goodfellow}\ \emph {et~al.}(2016)\citenamefont
  {Goodfellow}, \citenamefont {Bengio},\ and\ \citenamefont
  {Courville}}]{Goodfellow2016DeepLearning}%
  \BibitemOpen
  \bibfield  {author} {\bibinfo {author} {\bibfnamefont {I.}~\bibnamefont
  {Goodfellow}}, \bibinfo {author} {\bibfnamefont {Y.}~\bibnamefont {Bengio}},
  \ and\ \bibinfo {author} {\bibfnamefont {A.}~\bibnamefont {Courville}},\
  }\href {http://www.deeplearningbook.org} {\emph {\bibinfo {title} {{Deep
  Learning}}}}\ (\bibinfo  {publisher} {MIT Press},\ \bibinfo {address}
  {Cambridge},\ \bibinfo {year} {2016})\BibitemShut {NoStop}%
\bibitem [{\citenamefont {Becca}\ and\ \citenamefont
  {Sorella}(2017)}]{Becca2017QuantumSystems}%
  \BibitemOpen
  \bibfield  {author} {\bibinfo {author} {\bibfnamefont {F.}~\bibnamefont
  {Becca}}\ and\ \bibinfo {author} {\bibfnamefont {S.}~\bibnamefont
  {Sorella}},\ }\href {\doibase 10.1017/9781316417041} {\emph {\bibinfo {title}
  {{Quantum Monte Carlo Approaches for Correlated Systems}}}}\ (\bibinfo
  {publisher} {Cambridge University Press},\ \bibinfo {address} {Cambridge},\
  \bibinfo {year} {2017})\BibitemShut {NoStop}%
\bibitem [{\citenamefont {Hu}\ \emph {et~al.}(2013)\citenamefont {Hu},
  \citenamefont {Becca}, \citenamefont {Parola},\ and\ \citenamefont
  {Sorella}}]{Hu2013DirectAntiferromagnetism}%
  \BibitemOpen
  \bibfield  {author} {\bibinfo {author} {\bibfnamefont {W.-J.}\ \bibnamefont
  {Hu}}, \bibinfo {author} {\bibfnamefont {F.}~\bibnamefont {Becca}}, \bibinfo
  {author} {\bibfnamefont {A.}~\bibnamefont {Parola}}, \ and\ \bibinfo {author}
  {\bibfnamefont {S.}~\bibnamefont {Sorella}},\ }\href {\doibase
  10.1103/PhysRevB.88.060402} {\bibfield  {journal} {\bibinfo  {journal} {Phys.
  Rev. B}\ }\textbf {\bibinfo {volume} {88}},\ \bibinfo {pages} {060402(R)}
  (\bibinfo {year} {2013})}\BibitemShut {NoStop}%
\bibitem [{\citenamefont {Gong}\ \emph {et~al.}(2014)\citenamefont {Gong},
  \citenamefont {Zhu}, \citenamefont {Sheng}, \citenamefont {Motrunich},\ and\
  \citenamefont {Fisher}}]{Gong2014PlaquetteModel}%
  \BibitemOpen
  \bibfield  {author} {\bibinfo {author} {\bibfnamefont {S.-S.}\ \bibnamefont
  {Gong}}, \bibinfo {author} {\bibfnamefont {W.}~\bibnamefont {Zhu}}, \bibinfo
  {author} {\bibfnamefont {D.~N.}\ \bibnamefont {Sheng}}, \bibinfo {author}
  {\bibfnamefont {O.~I.}\ \bibnamefont {Motrunich}}, \ and\ \bibinfo {author}
  {\bibfnamefont {M.~P.~A.}\ \bibnamefont {Fisher}},\ }\href {\doibase
  10.1103/PhysRevLett.113.027201} {\bibfield  {journal} {\bibinfo  {journal}
  {Phys. Rev. Lett.}\ }\textbf {\bibinfo {volume} {113}},\ \bibinfo {pages}
  {027201} (\bibinfo {year} {2014})}\BibitemShut {NoStop}%
\bibitem [{\citenamefont {Zen}\ \emph {et~al.}(2020)\citenamefont {Zen},
  \citenamefont {My}, \citenamefont {Tan}, \citenamefont {Hebert},
  \citenamefont {Gattobigio}, \citenamefont {Miniatura}, \citenamefont
  {Poletti},\ and\ \citenamefont {Bressan}}]{Zen2020FindingStates}%
  \BibitemOpen
  \bibfield  {author} {\bibinfo {author} {\bibfnamefont {R.}~\bibnamefont
  {Zen}}, \bibinfo {author} {\bibfnamefont {L.}~\bibnamefont {My}}, \bibinfo
  {author} {\bibfnamefont {R.}~\bibnamefont {Tan}}, \bibinfo {author}
  {\bibfnamefont {F.}~\bibnamefont {Hebert}}, \bibinfo {author} {\bibfnamefont
  {M.}~\bibnamefont {Gattobigio}}, \bibinfo {author} {\bibfnamefont
  {C.}~\bibnamefont {Miniatura}}, \bibinfo {author} {\bibfnamefont
  {D.}~\bibnamefont {Poletti}}, \ and\ \bibinfo {author} {\bibfnamefont
  {S.}~\bibnamefont {Bressan}},\ }\href {http://arxiv.org/abs/2002.02618}
  {\bibfield  {journal} {\bibinfo  {journal} {arXiv:2002.02618}\ } (\bibinfo
  {year} {2020})}\BibitemShut {NoStop}%
\bibitem [{\citenamefont {Thibaut}\ \emph {et~al.}(2019)\citenamefont
  {Thibaut}, \citenamefont {Roscilde},\ and\ \citenamefont
  {Mezzacapo}}]{Thibaut2019Long-rangeLattice}%
  \BibitemOpen
  \bibfield  {author} {\bibinfo {author} {\bibfnamefont {J.}~\bibnamefont
  {Thibaut}}, \bibinfo {author} {\bibfnamefont {T.}~\bibnamefont {Roscilde}}, \
  and\ \bibinfo {author} {\bibfnamefont {F.}~\bibnamefont {Mezzacapo}},\ }\href
  {\doibase 10.1103/PhysRevB.100.155148} {\bibfield  {journal} {\bibinfo
  {journal} {Phys. Rev. B}\ }\textbf {\bibinfo {volume} {100}},\ \bibinfo
  {pages} {155148} (\bibinfo {year} {2019})}\BibitemShut {NoStop}%
\bibitem [{\citenamefont {Carleo}\ \emph
  {et~al.}(2019{\natexlab{b}})\citenamefont {Carleo}, \citenamefont {Choo},
  \citenamefont {Hofmann}, \citenamefont {Smith}, \citenamefont {Westerhout},
  \citenamefont {Alet}, \citenamefont {Davis}, \citenamefont {Efthymiou},
  \citenamefont {Glasser}, \citenamefont {Lin}, \citenamefont {Mauri},
  \citenamefont {Mazzola}, \citenamefont {Mendl}, \citenamefont {van
  Nieuwenburg}, \citenamefont {O’Reilly}, \citenamefont {Th{\'{e}}veniaut},
  \citenamefont {Torlai}, \citenamefont {Vicentini},\ and\ \citenamefont
  {Wietek}}]{Carleo2019NetKet:Systems}%
  \BibitemOpen
  \bibfield  {author} {\bibinfo {author} {\bibfnamefont {G.}~\bibnamefont
  {Carleo}}, \bibinfo {author} {\bibfnamefont {K.}~\bibnamefont {Choo}},
  \bibinfo {author} {\bibfnamefont {D.}~\bibnamefont {Hofmann}}, \bibinfo
  {author} {\bibfnamefont {J.~E.}\ \bibnamefont {Smith}}, \bibinfo {author}
  {\bibfnamefont {T.}~\bibnamefont {Westerhout}}, \bibinfo {author}
  {\bibfnamefont {F.}~\bibnamefont {Alet}}, \bibinfo {author} {\bibfnamefont
  {E.~J.}\ \bibnamefont {Davis}}, \bibinfo {author} {\bibfnamefont
  {S.}~\bibnamefont {Efthymiou}}, \bibinfo {author} {\bibfnamefont
  {I.}~\bibnamefont {Glasser}}, \bibinfo {author} {\bibfnamefont {S.-H.}\
  \bibnamefont {Lin}}, \bibinfo {author} {\bibfnamefont {M.}~\bibnamefont
  {Mauri}}, \bibinfo {author} {\bibfnamefont {G.}~\bibnamefont {Mazzola}},
  \bibinfo {author} {\bibfnamefont {C.~B.}\ \bibnamefont {Mendl}}, \bibinfo
  {author} {\bibfnamefont {E.~P.~L.}\ \bibnamefont {van Nieuwenburg}}, \bibinfo
  {author} {\bibfnamefont {O.}~\bibnamefont {O’Reilly}}, \bibinfo {author}
  {\bibfnamefont {H.}~\bibnamefont {Th{\'{e}}veniaut}}, \bibinfo {author}
  {\bibfnamefont {G.}~\bibnamefont {Torlai}}, \bibinfo {author} {\bibfnamefont
  {F.}~\bibnamefont {Vicentini}}, \ and\ \bibinfo {author} {\bibfnamefont
  {A.}~\bibnamefont {Wietek}},\ }\href {\doibase 10.1016/J.SOFTX.2019.100311}
  {\bibfield  {journal} {\bibinfo  {journal} {SoftwareX}\ }\textbf {\bibinfo
  {volume} {10}},\ \bibinfo {pages} {100311} (\bibinfo {year}
  {2019}{\natexlab{b}})}\BibitemShut {NoStop}%
\bibitem [{\citenamefont {Park}\ and\ \citenamefont
  {Kastoryano}(2020)}]{Park2020GeometryStates}%
  \BibitemOpen
  \bibfield  {author} {\bibinfo {author} {\bibfnamefont {C.-Y.}\ \bibnamefont
  {Park}}\ and\ \bibinfo {author} {\bibfnamefont {M.~J.}\ \bibnamefont
  {Kastoryano}},\ }\href {\doibase 10.1103/physrevresearch.2.023232} {\bibfield
   {journal} {\bibinfo  {journal} {Phys. Rev. Research}\ }\textbf {\bibinfo
  {volume} {2}},\ \bibinfo {pages} {023232} (\bibinfo {year}
  {2020})}\BibitemShut {NoStop}%
\bibitem [{\citenamefont {Mezzacapo}\ \emph {et~al.}(2009)\citenamefont
  {Mezzacapo}, \citenamefont {Schuch}, \citenamefont {Boninsegni},\ and\
  \citenamefont {Cirac}}]{Mezzacapo2009Ground-stateCarlo}%
  \BibitemOpen
  \bibfield  {author} {\bibinfo {author} {\bibfnamefont {F.}~\bibnamefont
  {Mezzacapo}}, \bibinfo {author} {\bibfnamefont {N.}~\bibnamefont {Schuch}},
  \bibinfo {author} {\bibfnamefont {M.}~\bibnamefont {Boninsegni}}, \ and\
  \bibinfo {author} {\bibfnamefont {J.~I.}\ \bibnamefont {Cirac}},\ }\href
  {\doibase 10.1088/1367-2630/11/8/083026} {\bibfield  {journal} {\bibinfo
  {journal} {New J. Phys.}\ }\textbf {\bibinfo {volume} {11}},\ \bibinfo
  {pages} {083026} (\bibinfo {year} {2009})}\BibitemShut {NoStop}%
\bibitem [{\citenamefont {Ferrari}\ \emph {et~al.}(2019)\citenamefont
  {Ferrari}, \citenamefont {Becca},\ and\ \citenamefont
  {Carrasquilla}}]{Ferrari2019NeuralFunctions}%
  \BibitemOpen
  \bibfield  {author} {\bibinfo {author} {\bibfnamefont {F.}~\bibnamefont
  {Ferrari}}, \bibinfo {author} {\bibfnamefont {F.}~\bibnamefont {Becca}}, \
  and\ \bibinfo {author} {\bibfnamefont {J.}~\bibnamefont {Carrasquilla}},\
  }\href {\doibase 10.1103/PhysRevB.100.125131} {\bibfield  {journal} {\bibinfo
   {journal} {Phys. Rev. B}\ }\textbf {\bibinfo {volume} {100}},\ \bibinfo
  {pages} {125131} (\bibinfo {year} {2019})}\BibitemShut {NoStop}%
\bibitem [{\citenamefont {Troyer}\ and\ \citenamefont
  {Wiese}(2005)}]{Troyer2005ComputationalSimulations}%
  \BibitemOpen
  \bibfield  {author} {\bibinfo {author} {\bibfnamefont {M.}~\bibnamefont
  {Troyer}}\ and\ \bibinfo {author} {\bibfnamefont {U.~J.}\ \bibnamefont
  {Wiese}},\ }\href {\doibase 10.1103/PhysRevLett.94.170201} {\bibfield
  {journal} {\bibinfo  {journal} {Phys. Rev. Lett.}\ }\textbf {\bibinfo
  {volume} {94}},\ \bibinfo {pages} {170201} (\bibinfo {year}
  {2005})}\BibitemShut {NoStop}%
\bibitem [{\citenamefont {Richter}\ \emph {et~al.}(1994)\citenamefont
  {Richter}, \citenamefont {Ivanov},\ and\ \citenamefont
  {Retzlaff}}]{Richter1994OnAntiferromagnet}%
  \BibitemOpen
  \bibfield  {author} {\bibinfo {author} {\bibfnamefont {J.}~\bibnamefont
  {Richter}}, \bibinfo {author} {\bibfnamefont {N.~B.}\ \bibnamefont {Ivanov}},
  \ and\ \bibinfo {author} {\bibfnamefont {K.}~\bibnamefont {Retzlaff}},\
  }\href {\doibase 10.1209/0295-5075/25/7/012} {\bibfield  {journal} {\bibinfo
  {journal} {Europhys. Lett.}\ }\textbf {\bibinfo {volume} {25}},\ \bibinfo
  {pages} {545} (\bibinfo {year} {1994})}\BibitemShut {NoStop}%
\bibitem [{\citenamefont {Voigt}\ \emph {et~al.}(1997)\citenamefont {Voigt},
  \citenamefont {Richter},\ and\ \citenamefont
  {Ivanov}}]{Voigt1997Marshall-PeierlsAntiferromagnet}%
  \BibitemOpen
  \bibfield  {author} {\bibinfo {author} {\bibfnamefont {A.}~\bibnamefont
  {Voigt}}, \bibinfo {author} {\bibfnamefont {J.}~\bibnamefont {Richter}}, \
  and\ \bibinfo {author} {\bibfnamefont {N.~B.}\ \bibnamefont {Ivanov}},\
  }\href {\doibase 10.1016/S0378-4371(97)00330-0} {\bibfield  {journal}
  {\bibinfo  {journal} {Physica A}\ }\textbf {\bibinfo {volume} {245}},\
  \bibinfo {pages} {269} (\bibinfo {year} {1997})}\BibitemShut {NoStop}%
\bibitem [{\citenamefont {Sehayek}\ \emph {et~al.}(2019)\citenamefont
  {Sehayek}, \citenamefont {Golubeva}, \citenamefont {Albergo}, \citenamefont
  {Kulchytskyy}, \citenamefont {Torlai},\ and\ \citenamefont
  {Melko}}]{Sehayek2019LearnabilityMachines}%
  \BibitemOpen
  \bibfield  {author} {\bibinfo {author} {\bibfnamefont {D.}~\bibnamefont
  {Sehayek}}, \bibinfo {author} {\bibfnamefont {A.}~\bibnamefont {Golubeva}},
  \bibinfo {author} {\bibfnamefont {M.~S.}\ \bibnamefont {Albergo}}, \bibinfo
  {author} {\bibfnamefont {B.}~\bibnamefont {Kulchytskyy}}, \bibinfo {author}
  {\bibfnamefont {G.}~\bibnamefont {Torlai}}, \ and\ \bibinfo {author}
  {\bibfnamefont {R.~G.}\ \bibnamefont {Melko}},\ }\href {\doibase
  10.1103/PhysRevB.100.195125} {\bibfield  {journal} {\bibinfo  {journal}
  {Phys. Rev. B}\ }\textbf {\bibinfo {volume} {100}},\ \bibinfo {pages}
  {195125} (\bibinfo {year} {2019})}\BibitemShut {NoStop}%
\bibitem [{\citenamefont {Lin}\ \emph {et~al.}(2017)\citenamefont {Lin},
  \citenamefont {Tegmark},\ and\ \citenamefont {Rolnick}}]{Lin2017WhyWell}%
  \BibitemOpen
  \bibfield  {author} {\bibinfo {author} {\bibfnamefont {H.~W.}\ \bibnamefont
  {Lin}}, \bibinfo {author} {\bibfnamefont {M.}~\bibnamefont {Tegmark}}, \ and\
  \bibinfo {author} {\bibfnamefont {D.}~\bibnamefont {Rolnick}},\ }\href
  {\doibase 10.1007/s10955-017-1836-5} {\bibfield  {journal} {\bibinfo
  {journal} {J. Stat. Phys.}\ }\textbf {\bibinfo {volume} {168}},\ \bibinfo
  {pages} {1223} (\bibinfo {year} {2017})}\BibitemShut {NoStop}%
\bibitem [{\citenamefont {Valle-P{\'{e}}rez}\ \emph {et~al.}(2018)\citenamefont
  {Valle-P{\'{e}}rez}, \citenamefont {Camargo},\ and\ \citenamefont
  {Louis}}]{Valle-Perez2018DeepFunctions}%
  \BibitemOpen
  \bibfield  {author} {\bibinfo {author} {\bibfnamefont {G.}~\bibnamefont
  {Valle-P{\'{e}}rez}}, \bibinfo {author} {\bibfnamefont {C.~Q.}\ \bibnamefont
  {Camargo}}, \ and\ \bibinfo {author} {\bibfnamefont {A.~A.}\ \bibnamefont
  {Louis}},\ }\href {http://arxiv.org/abs/1805.08522} {\bibfield  {journal}
  {\bibinfo  {journal} {arXiv:1805.08522}\ } (\bibinfo {year}
  {2018})}\BibitemShut {NoStop}%
\bibitem [{\citenamefont {De~Palma}\ \emph {et~al.}(2018)\citenamefont
  {De~Palma}, \citenamefont {Kiani},\ and\ \citenamefont
  {Lloyd}}]{DePalma2018RandomFunctions}%
  \BibitemOpen
  \bibfield  {author} {\bibinfo {author} {\bibfnamefont {G.}~\bibnamefont
  {De~Palma}}, \bibinfo {author} {\bibfnamefont {B.~T.}\ \bibnamefont {Kiani}},
  \ and\ \bibinfo {author} {\bibfnamefont {S.}~\bibnamefont {Lloyd}},\ }\href
  {http://arxiv.org/abs/1812.10156} {\bibfield  {journal} {\bibinfo  {journal}
  {arXiv:1812.10156}\ } (\bibinfo {year} {2018})}\BibitemShut {NoStop}%
\bibitem [{\citenamefont {Kovesi}(2015)}]{Kovesi2015GoodThem}%
  \BibitemOpen
  \bibfield  {author} {\bibinfo {author} {\bibfnamefont {P.}~\bibnamefont
  {Kovesi}},\ }\href {http://arxiv.org/abs/1509.03700} {\bibfield  {journal}
  {\bibinfo  {journal} {arXiv:1509.03700}\ } (\bibinfo {year}
  {2015})}\BibitemShut {NoStop}%
\bibitem [{\citenamefont {Schollw{\"{o}}ck}(2005)}]{Schollwock2005TheGroup}%
  \BibitemOpen
  \bibfield  {author} {\bibinfo {author} {\bibfnamefont {U.}~\bibnamefont
  {Schollw{\"{o}}ck}},\ }\href {\doibase 10.1103/RevModPhys.77.259} {\bibfield
  {journal} {\bibinfo  {journal} {Rev. Mod. Phys.}\ }\textbf {\bibinfo {volume}
  {77}},\ \bibinfo {pages} {259} (\bibinfo {year} {2005})}\BibitemShut
  {NoStop}%
\bibitem [{\citenamefont {Vieijra}\ \emph {et~al.}(2020)\citenamefont
  {Vieijra}, \citenamefont {Casert}, \citenamefont {Nys}, \citenamefont
  {De~Neve}, \citenamefont {Haegeman}, \citenamefont {Ryckebusch},\ and\
  \citenamefont {Verstraete}}]{Vieijra2020RestrictedSymmetries}%
  \BibitemOpen
  \bibfield  {author} {\bibinfo {author} {\bibfnamefont {T.}~\bibnamefont
  {Vieijra}}, \bibinfo {author} {\bibfnamefont {C.}~\bibnamefont {Casert}},
  \bibinfo {author} {\bibfnamefont {J.}~\bibnamefont {Nys}}, \bibinfo {author}
  {\bibfnamefont {W.}~\bibnamefont {De~Neve}}, \bibinfo {author} {\bibfnamefont
  {J.}~\bibnamefont {Haegeman}}, \bibinfo {author} {\bibfnamefont
  {J.}~\bibnamefont {Ryckebusch}}, \ and\ \bibinfo {author} {\bibfnamefont
  {F.}~\bibnamefont {Verstraete}},\ }\href {\doibase
  10.1103/PhysRevLett.124.097201} {\bibfield  {journal} {\bibinfo  {journal}
  {Phys. Rev. Lett.}\ }\textbf {\bibinfo {volume} {124}},\ \bibinfo {pages}
  {097201} (\bibinfo {year} {2020})}\BibitemShut {NoStop}%
\bibitem [{\citenamefont {Sorella}(2001)}]{Sorella2001GeneralizedCarlo}%
  \BibitemOpen
  \bibfield  {author} {\bibinfo {author} {\bibfnamefont {S.}~\bibnamefont
  {Sorella}},\ }\href {\doibase 10.1103/PhysRevB.64.024512} {\bibfield
  {journal} {\bibinfo  {journal} {Phys. Rev. B}\ }\textbf {\bibinfo {volume}
  {64}},\ \bibinfo {pages} {024512} (\bibinfo {year} {2001})}\BibitemShut
  {NoStop}%
\bibitem [{\citenamefont {He}\ \emph {et~al.}(2015)\citenamefont {He},
  \citenamefont {Zhang}, \citenamefont {Ren},\ and\ \citenamefont
  {Sun}}]{He2015DelvingClassification}%
  \BibitemOpen
  \bibfield  {author} {\bibinfo {author} {\bibfnamefont {K.}~\bibnamefont
  {He}}, \bibinfo {author} {\bibfnamefont {X.}~\bibnamefont {Zhang}}, \bibinfo
  {author} {\bibfnamefont {S.}~\bibnamefont {Ren}}, \ and\ \bibinfo {author}
  {\bibfnamefont {J.}~\bibnamefont {Sun}},\ }\href
  {http://arxiv.org/abs/1502.01852} {\bibfield  {journal} {\bibinfo  {journal}
  {arXiv:1502.01852}\ } (\bibinfo {year} {2015})}\BibitemShut {NoStop}%
\bibitem [{\citenamefont {Heine}(1960)}]{Heine1960GroupMechanics}%
  \BibitemOpen
  \bibfield  {author} {\bibinfo {author} {\bibfnamefont {V.}~\bibnamefont
  {Heine}},\ }\href@noop {} {\emph {\bibinfo {title} {{Group Theory in Quantum
  Mechanics}}}},\ \bibinfo {series} {International Series in Natural
  Philosophy}, Vol.~\bibinfo {volume} {91}\ (\bibinfo  {publisher} {Pergamon},\
  \bibinfo {address} {Oxford},\ \bibinfo {year} {1960})\BibitemShut {NoStop}%
\bibitem [{\citenamefont {Virtanen}\ \emph {et~al.}(2020)\citenamefont
  {Virtanen}, \citenamefont {Gommers}, \citenamefont {Oliphant}, \citenamefont
  {Haberland}, \citenamefont {Reddy}, \citenamefont {Cournapeau}, \citenamefont
  {Burovski}, \citenamefont {Peterson}, \citenamefont {Weckesser},
  \citenamefont {Bright}, \citenamefont {van~der Walt}, \citenamefont {Brett},
  \citenamefont {Wilson}, \citenamefont {Millman}, \citenamefont {Mayorov},
  \citenamefont {Nelson}, \citenamefont {Jones}, \citenamefont {Kern},
  \citenamefont {Larson}, \citenamefont {Carey}, \citenamefont {Polat},
  \citenamefont {Feng}, \citenamefont {Moore}, \citenamefont {VanderPlas},
  \citenamefont {Laxalde}, \citenamefont {Perktold}, \citenamefont {Cimrman},
  \citenamefont {Henriksen}, \citenamefont {Quintero}, \citenamefont {Harris},
  \citenamefont {Archibald}, \citenamefont {Ribeiro}, \citenamefont
  {Pedregosa}, \citenamefont {van Mulbregt},\ and\ \citenamefont {{SciPy 1.0
  Contributors}}}]{Virtanen2020SciPyPython}%
  \BibitemOpen
  \bibfield  {author} {\bibinfo {author} {\bibfnamefont {P.}~\bibnamefont
  {Virtanen}}, \bibinfo {author} {\bibfnamefont {R.}~\bibnamefont {Gommers}},
  \bibinfo {author} {\bibfnamefont {T.~E.}\ \bibnamefont {Oliphant}}, \bibinfo
  {author} {\bibfnamefont {M.}~\bibnamefont {Haberland}}, \bibinfo {author}
  {\bibfnamefont {T.}~\bibnamefont {Reddy}}, \bibinfo {author} {\bibfnamefont
  {D.}~\bibnamefont {Cournapeau}}, \bibinfo {author} {\bibfnamefont
  {E.}~\bibnamefont {Burovski}}, \bibinfo {author} {\bibfnamefont
  {P.}~\bibnamefont {Peterson}}, \bibinfo {author} {\bibfnamefont
  {W.}~\bibnamefont {Weckesser}}, \bibinfo {author} {\bibfnamefont
  {J.}~\bibnamefont {Bright}}, \bibinfo {author} {\bibfnamefont {S.~J.}\
  \bibnamefont {van~der Walt}}, \bibinfo {author} {\bibfnamefont
  {M.}~\bibnamefont {Brett}}, \bibinfo {author} {\bibfnamefont
  {J.}~\bibnamefont {Wilson}}, \bibinfo {author} {\bibfnamefont {K.~J.}\
  \bibnamefont {Millman}}, \bibinfo {author} {\bibfnamefont {N.}~\bibnamefont
  {Mayorov}}, \bibinfo {author} {\bibfnamefont {A.~R.~J.}\ \bibnamefont
  {Nelson}}, \bibinfo {author} {\bibfnamefont {E.}~\bibnamefont {Jones}},
  \bibinfo {author} {\bibfnamefont {R.}~\bibnamefont {Kern}}, \bibinfo {author}
  {\bibfnamefont {E.}~\bibnamefont {Larson}}, \bibinfo {author} {\bibfnamefont
  {C.~J.}\ \bibnamefont {Carey}}, \bibinfo {author} {\bibfnamefont
  {I.}~\bibnamefont {Polat}}, \bibinfo {author} {\bibfnamefont
  {Y.}~\bibnamefont {Feng}}, \bibinfo {author} {\bibfnamefont {E.~W.}\
  \bibnamefont {Moore}}, \bibinfo {author} {\bibfnamefont {J.}~\bibnamefont
  {VanderPlas}}, \bibinfo {author} {\bibfnamefont {D.}~\bibnamefont {Laxalde}},
  \bibinfo {author} {\bibfnamefont {J.}~\bibnamefont {Perktold}}, \bibinfo
  {author} {\bibfnamefont {R.}~\bibnamefont {Cimrman}}, \bibinfo {author}
  {\bibfnamefont {I.}~\bibnamefont {Henriksen}}, \bibinfo {author}
  {\bibfnamefont {E.~A.}\ \bibnamefont {Quintero}}, \bibinfo {author}
  {\bibfnamefont {C.~R.}\ \bibnamefont {Harris}}, \bibinfo {author}
  {\bibfnamefont {A.~M.}\ \bibnamefont {Archibald}}, \bibinfo {author}
  {\bibfnamefont {A.~H.}\ \bibnamefont {Ribeiro}}, \bibinfo {author}
  {\bibfnamefont {F.}~\bibnamefont {Pedregosa}}, \bibinfo {author}
  {\bibfnamefont {P.}~\bibnamefont {van Mulbregt}}, \ and\ \bibinfo {author}
  {\bibnamefont {{SciPy 1.0 Contributors}}},\ }\href {\doibase
  10.1038/s41592-019-0686-2} {\bibfield  {journal} {\bibinfo  {journal} {Nat.
  Methods}\ }\textbf {\bibinfo {volume} {17}},\ \bibinfo {pages} {261}
  (\bibinfo {year} {2020})}\BibitemShut {NoStop}%
\bibitem [{\citenamefont {Baldassi}\ \emph {et~al.}(2020)\citenamefont
  {Baldassi}, \citenamefont {Pittorino},\ and\ \citenamefont
  {Zecchina}}]{Baldassi2020ShapingMinima}%
  \BibitemOpen
  \bibfield  {author} {\bibinfo {author} {\bibfnamefont {C.}~\bibnamefont
  {Baldassi}}, \bibinfo {author} {\bibfnamefont {F.}~\bibnamefont {Pittorino}},
  \ and\ \bibinfo {author} {\bibfnamefont {R.}~\bibnamefont {Zecchina}},\
  }\href {\doibase 10.1073/pnas.1908636117} {\bibfield  {journal} {\bibinfo
  {journal} {Proc. Natl. Acad. Sci.}\ }\textbf {\bibinfo {volume} {117}},\
  \bibinfo {pages} {161} (\bibinfo {year} {2020})}\BibitemShut {NoStop}%
\end{thebibliography}%
\end{document}